\documentclass[iop,twocolumn]{aastex631}
\usepackage{CJK}
\usepackage{mathrsfs}
\usepackage{amsmath}
\usepackage{float}
\usepackage{amstext}
\usepackage{soul}

\hypersetup{
   colorlinks,
   linkcolor={blue!88!black!80},
   citecolor={blue!88!black!80},
   urlcolor={blue!88!black!80}}

\newcommand{\ha}{H$\alpha$}
\newcommand{\hb}{H$\beta$}
\newcommand{\kms}{km s$^{-1}$}

\shorttitle{Dust Reverberation Mapping of IMBHs}

\turnoffeditone

\begin{document}
\begin{CJK}{UTF8}{gbsn}

\title{The Intermediate-Mass Black Hole Reverberation Mapping Project: \\ First Detection of Mid-Infrared Lags in Prototypical IMBHs in NGC 4395 and POX 52}

\correspondingauthor{Jingbo Sun, Hengxiao Guo }
\email{sunjingbo@shao.ac.cn (JBS)\\ hengxiaoguo@gmail.com (HXG)} 

\author[0000-0003-3472-4392]{Jingbo Sun} 
\affiliation{Shanghai Astronomical Observatory, Chinese Academy of Sciences, 80 Nandan Road, Shanghai 200030, People's Republic of China}
\affiliation{University of Chinese Academy of Sciences, 19A Yuquan Road, 100049, Beijing, People's Republic of China}

\author[0000-0001-8416-7059]{Hengxiao Guo} 
\affiliation{Shanghai Astronomical Observatory, Chinese Academy of Sciences, 80 Nandan Road, Shanghai 200030, People's Republic of China}

\author[0000-0002-4521-6281]{Wenwen Zuo}
\affiliation{Shanghai Astronomical Observatory, Chinese Academy of Sciences, 80 Nandan Road, Shanghai 200030, People's Republic of China}

\author[0000-0003-1523-9164]{Paulina Lira}
\affiliation{Departamento de Astronom \`{\i}a, Universidad de Chile, Casilla 36D, Santiago, Chile}

\author[0000-0002-4455-6946]{Minfeng Gu}
\affiliation{Shanghai Astronomical Observatory, Chinese Academy of Sciences, 80 Nandan Road, Shanghai 200030, People's Republic of China}

\author[0000-0002-8186-4753]{Philip G. Edwards} 
\affiliation{CSIRO Astronomy and Space Science, PO Box 76, Epping, NSW, 1710, Australia}

\author[0000-0002-2052-6400]{Shu Wang}
\affiliation{Department of Physics \& Astronomy, Seoul National University, Seoul 08826, Republic of Korea}

\author[0000-0002-5841-3348]{Jamie Stevens}
\affiliation{CSIRO Astronomy and Space Science, PO Box 76, Epping, NSW, 1710, Australia}

\author[0000-0003-4341-0029]{Tao An}
\affiliation{Shanghai Astronomical Observatory, Chinese Academy of Sciences, 80 Nandan Road, Shanghai 200030, People's Republic of China} 

\author[0000-0002-5248-2422]{Samuzal Barua}
\affiliation{Shanghai Astronomical Observatory, Chinese Academy of Sciences, 80 Nandan Road, Shanghai 200030, People's Republic of China}

\author[0000-0002-4223-2198]{Zhen-yi Cai}
\affiliation{Department of Astronomy, University of Science and Technology of China, Hefei, Anhui 230026, People's Republic of China} 
\affiliation{School of Astronomy and Space Science, University of Science and Technology of China, Hefei 230026, People's Republic of China} 

\author[0000-0002-1530-2680]{Haicheng Feng}
\affiliation{Yunnan Observatories, Chinese Academy of Sciences, 396 Yangfangwang, Guandu District, Kunming 650216, Yunnan, People's Republic of China}
\affiliation{Key Laboratory for the Structure and Evolution of Celestial Objects, Chinese Academy of Sciences, Kunming 650216, Yunnan,
People's Republic of China}
\affiliation{Center for Astronomical Mega-Science, Chinese Academy of Sciences, 20A Datun Road, Chaoyang District, Beijing 100012,
People's Republic of China}

\author[0000-0002-9331-4388]{Alok C. Gupta}
\affiliation{Aryabhatta Research Institute of Observational Sciences (ARIES), Manora Peak, Nainital 263001, India}

\author[0000-0001-6947-5846]{Luis C. Ho}
\affiliation{Department of Astronomy, School of Physics, Peking University, Beijing 100871, People's Republic of China} 
\affiliation{Kavli Institute for Astronomy and Astrophysics, Peking University, Beijing, 100871, People's Republic of China}

\author[0000-0002-1134-4015]{Dragana Ili\'{c}}
\affiliation{University of Belgrade - Faculty of Mathematics, Department of Astronomy, Studentski trg 16, Belgrade, Serbia}
\affiliation{Hamburger Sternwarte, Universitat Hamburg, Gojenbergsweg 112, D-21029 Hamburg, Germany}

\author[0000-0001-5139-1978]{Andjelka B. Kova\v{c}evi\'{c}}
\affiliation{University of Belgrade - Faculty of Mathematics, Department of Astronomy, Studentski trg 16, Belgrade, Serbia}

\author[0000-0003-3823-3419]{ShaSha Li}
\affiliation{Yunnan Observatories, Chinese Academy of Sciences, 396 Yangfangwang, Guandu District, Kunming 650216, Yunnan, People's Republic of China}
\affiliation{Key Laboratory for the Structure and Evolution of Celestial Objects, Chinese Academy of Sciences, Kunming 650216, Yunnan,
People's Republic of China}
\affiliation{Center for Astronomical Mega-Science, Chinese Academy of Sciences, 20A Datun Road, Chaoyang District, Beijing 100012,
People's Republic of China}

\author[0000-0003-4440-259X]{Mar Mezcua}
\affiliation{Institute of Space Sciences (ICE, CSIC), Campus UAB, Carrer de Magrans, 08193 Barcelona, Spain}
\affiliation{Institut d'Estudis Espacials de Catalunya (IEEC), Edifici RDIT, Campus UPC, 08860 Castelldefels (Barcelona), Spain}

\author[0000-0003-2398-7664]{Luka \v{C}. Popovi\'{c}}
\affiliation{University of Belgrade - Faculty of Mathematics, Department of Astronomy, Studentski trg 16, Belgrade, Serbia}
\affiliation{Astronomical Observatory, Volgina 7, 11060 Belgrade, Serbia}

\author[0000-0003-0820-4692]{Paula S\'{a}nchez-S\'{a}ez}
\affiliation{European Southern Observatory, Karl-Schwarzschild-Str. 2, 85748, Garching, Germany}

\author[0000-0002-0771-2153]{Mouyuan Sun}
\affiliation{Department of Astronomy, Xiamen University, Xiamen, Fujian 361005, People's Republic of China} 

\author[0000-0001-5012-2362]{Rongfeng Shen}
\affiliation{School of Physics and Astronomy, Sun Yat-Sen University, Zhuhai, 519082, People's Republic of China}
\affiliation{CSST Science Center for the Guangdong-Hongkong-Macau Greater Bay Area, Sun Yat-Sen University, Zhuhai, 519082, People's Republic of China}

\author[0000-0002-1912-0024]{Vivian U}
\affiliation{IPAC, Caltech, 1200 E. California Blvd., Pasadena, CA 91125, USA}
\affiliation{Department of Physics and Astronomy, 4129 Frederick Reines Hall, University of California, Irvine, CA 92697, USA}

\author[0009-0008-5761-3701]{Oliver Vince}
\affiliation{Astronomical Observatory, Volgina 7, 11060 Belgrade, Serbia}
\affiliation{Shanghai Astronomical Observatory, Chinese Academy of Sciences, 80 Nandan Road, Shanghai 200030, People's Republic of China}

\author[0000-0002-4419-6434]{Junxian Wang}
\affiliation{CAS Key Laboratory for Research in Galaxies and Cosmology, Department of Astronomy, University of Science and Technology of China, Hefei, Anhui 230026, People's Republic of China} 
\affiliation{School of Astronomy and Space Science, University of Science and Technology of China, Hefei 230026, People's Republic of China}

\author[0000-0002-7350-6913]{Xuebing Wu}
\affiliation{Department of Astronomy, School of Physics, Peking University, Beijing 100871, People's Republic of China} 
\affiliation{Kavli Institute for Astronomy and Astrophysics, Peking University, Beijing, 100871, People's Republic of China} 

\author[0000-0003-0644-9282]{Zhefu Yu}
\affiliation{Kavli Institute for Particle Astrophysics and Cosmology (KIPAC), Stanford University, Stanford CA 94305, USA}

\author[0000-0002-9634-2923]{Zhenya Zheng}
\affiliation{Shanghai Astronomical Observatory, Chinese Academy of Sciences, 80 Nandan Road, Shanghai 200030, People's Republic of China}

\begin{abstract}
The search for robust evidence of intermediate-mass black holes (IMBHs) is crucial for understanding black hole seeding process and the formation of supermassive black holes in the early Universe. NGC~4395 and POX~52 are two prototypical IMBH hosts, both exhibiting multi-line evidence of low-mass black hole activity. Here, we report the first detection of mid-infrared (MIR) lags in response to optical variability, with measurements of $3.0^{+2.4}_{-1.9}$ days for NGC 4395 and $35.2^{+14.2}_{-11.7}$ days for POX~52 at $3.4$ $\mu$m, respectively, using archival optical data and observations from the Wide-field Infrared Survey Explorer (WISE). This detection provides the first reverberation evidence of low-mass black hole activity in POX 52. The time lags of these two low-mass, low-luminosity active galactic nuclei (AGNs) generally follow the extent of the $R_{\rm dust}-L_{\rm 5100}$ relation found in higher-mass AGNs. Based on an empirical relation between the broad-line region and dusty torus size, we constrain the black hole mass of POX 52 to log($M_{\rm BH}$/$M_\odot$) = 5.5 $\pm$ 0.37 (systemic and statistical errors), confirming its IMBH nature. Furthermore, long-term optical continuum monitoring of POX 52 reveals a mild inter-band lag of $\lesssim$ 1 day. However, no significant intranight variability was detected during its one-night, high-cadence monitoring, which we attribute to the longer duty cycle of fast variability in POX 52 compared to that in NGC 4395.

\end{abstract}


\section{Introduction}  
\label{sec:intro}
Accurate mass measurements of intermediate-mass black holes (IMBHs, $M_{\rm BH} \sim 10^{2}$ to $10^{6} M_{\odot}$) and a comprehensive census are crucial for unraveling essential clues about the seeding mechanisms and formation of supermassive black holes (SMBHs) during the early Universe \citep[see reviews,][]{Mezcua17, Greene2020, Inayoshi20, Volonteri21, Reines22}. At least three hosts or counterparts for IMBHs have been proposed: globular clusters (e.g., 47 Tuc, \citealp{Kiziltan17}; G1, \citealp{Gebhardt05}), ultra-/hyper-luminous X-ray sources (ULX/HLX) \citep[e.g., HLX-1,][]{Webb12}, and the nuclei of galaxies. Given the limited number of IMBH candidates in globular clusters and ULX/HLX, more attention has been drawn to IMBHs residing in galaxies \citep[e.g.,][]{Greene04, Greene07, Dong12, Liu18, Chilingarian18, LiuWJ25}, especially those in low-mass dwarf galaxies \citep[e.g.,][]{Reines13, Pucha25}. Typically, relatively reliable candidates in dwarf galaxies are selected via single-epoch spectra, where the black hole (BH) mass ($\lesssim 10^6 M_{\odot}$) is estimated from the broad emission lines originating from the broad line region (such as $\rm H\alpha$ or $\rm H\beta$). However, temporary stellar processes, such as supernova remnants, can mimic the weak broad line signature used for mass estimation. In addition, single-epoch BH mass measurements are subject to significant uncertainties \citep[$\sim$0.5 dex; e.g.,][]{Shen13}. Therefore, to conclusively ascertain the IMBH nature and facilitate future statistical studies, precise measurements of IMBH masses are essential.

Dynamical modeling of stellar or gas motion provides a robust approach for estimating SMBH masses \citep{Kormendy95,Kormendy13}. However, extending this approach to the lower-mass regime presents significant challenges. These include the insufficient instrumental capability to resolve the BH's sphere of influence, a more complex stellar mass-to-light ratio in low-mass galaxies due to star formation and dust, and the presence of non-axisymmetric structures in the galactic center \citep{Greene20}. Thus, this dynamical method has only been applied to a few very nearby galaxies hosting an inactive IMBH, \citep[e.g., NGC 205,][]{Nguyen19}. Among them, NGC 4395 stands out as the only robust broad-line active galactic nucleus (AGN) hosting an IMBH confirmed by the dynamical method \citep{denBrok15, Brum19}. 

Broad-line reverberation mapping (RM) is an alternative technique for directly estimating BH masses \citep{Blandford82}, applicable across a wide range of AGN masses. Broad-line RM involves measuring the time delay between variations in the continuum emission from the accretion disk and the corresponding response in the broad emission lines. This delay is used to determine the size of the broad-line region (BLR), and, when combined with the gas velocity derived from the linewidth, allows for the estimation of the BH mass from SMBHs \citep{Kaspi00,Bentz09,Bentz2013,Du14,Barth15, Lira18,U22,Yu23,Woo24,Shen24} to the IMBHs \citep[e.g., NGC 4395,][]{Peterson05, Woo19, Cho21, Pandey24}.

Additionally, continuum RM (CRM) offers a distinct approach to mapping accretion disk sizes by leveraging the wavelength-dependent lags predicted by the standard thin disk model \citep{Shakura1973}, in which the time lags are expected to scale with the wavelength as $\tau \sim \lambda^{4/3}$. CRM facilitates the measurements of time delays between different UV/optical wavelengths of continuum emission, reprocessed from hard X-rays emitted by the corona above the central region \citep[i.e., the lamp-post reprocessing model,][]{Krolik1991,Cackett2007}, allowing for comparison with theoretical predictions. This technique is primarily utilized to explore AGN variability mechanisms and the validity of standard thin disk and lamp-post models \citep[e.g.,][]{Cai18,Edelson2019}. Recently, \citet{Wang23} found that the BLR size of H$\beta$ is consistently $\sim$8 times that of the optical continuum emitting region at 5100 \AA, extending from SMBHs to the IMBH regime. This not only shows that continuum luminosity scales with the size of different AGN structures, e.g., accretion disk, BLR, and torus, but also demonstrates the potential applicability of CRM for estimating BH masses across a broad range without requiring expensive spectroscopic monitoring and reaching much fainter magnitudes.

In the same spirit, the dusty torus size is also expected to scale with the continuum luminosity, hence the BLR size, as the dust sublimation radius is determined by a temperature threshold governed by the luminosity. Deciphering the lag information between UV/optical emissions from the accretion disk and infrared (IR) emissions from the dusty torus, which are reprocessed from the UV/optical emissions, also provide strong constrain for BH mass. Reliable IR lag measurements can be traced back to \cite{Clavel89}, which showed a $\sim$400-day lag in Fairall 9 at near-IR (NIR) wavelengths ($\sim$2 $\mu$m, $K$-band) relative to UV/optical wavelengths. Later on, $K$-band lags are detected in a variety of individual AGNs, such as NGC 4151 \citep{Penston71, Oknyanskii93, Oknyanskij99, Minezaki04, Lyu2021}, NGC 3783 \citep{Glass92,Lira11}, NGC 1566 \citep{Baribaud92}, Mrk 744 \citep{Nelson96}, MR 2251-178 \citep{Lira11} and MCG-6-30-15 \citep{Lira15}. 

A major advancement came from the ground-based Multicolor Active Galactic Nuclei Monitoring (MAGNUM) project \citep{Yoshii2002, Yoshii2003}, which is operated from 2001 to 2008 and measured NIR lags for 36 type 1 AGNs. These measurements established a robust dust radius-luminosity relation, with torus sizes typically $4$–$5$ times larger than the BLR radii inferred from broad Balmer lines \citep{Minezaki19}. More recently, \citet{Lira24} used the UltraVISTA survey \citep{UltraVISTA12} to measure NIR time lags for 13 AGNs at redshifts between 0.3 and 0.8, finding consistency with the local radius–luminosity relation after applying K-corrections.

The advent of the space-based Wide-field Infrared Survey Explorer (WISE) has further expanded these efforts, facilitating mid-infrared (MIR) lag measurements at 3.4 $\mu$m (W1) and 4.6 $\mu$m (W2) for thousands of AGNs \citep{Lyu19, Yang20, Chen23, Mandal24}. However, it is important to note that while these studies have provided extensive data on SMBHs, no dust RM studies has been employed for IMBHs, except for a marginal lag detection at the NIR band in NGC 4395 \citep{Minezaki06}.

In this Letter, we report the first detection of MIR reverberation signals from two prototypical IMBHs in NGC 4395 and POX 52. Using the measured MIR lags, we are able to constrain the BH mass of POX 52 and validate its IMBH nature. This paper is organized as follows: Section \ref{sec:obs} describes the RM project, target properties, observations, and data analysis. Section \ref{sec:results} presents the results, and Section \ref{sec:diss} discusses their implications. We conclude in Section \ref{sec:conclusion}. Throughout this paper, we use the $\Lambda$CDM cosmology, with $H_{\rm 0}$ = 70.0 $\rm km~ s^{-1}~ Mpc^{-1}$, and $\Omega_{\rm m}$ = 0.3.

\section{Observations and Data Analysis}\label{sec:obs}

\begin{figure*}
    \includegraphics[width=\linewidth]{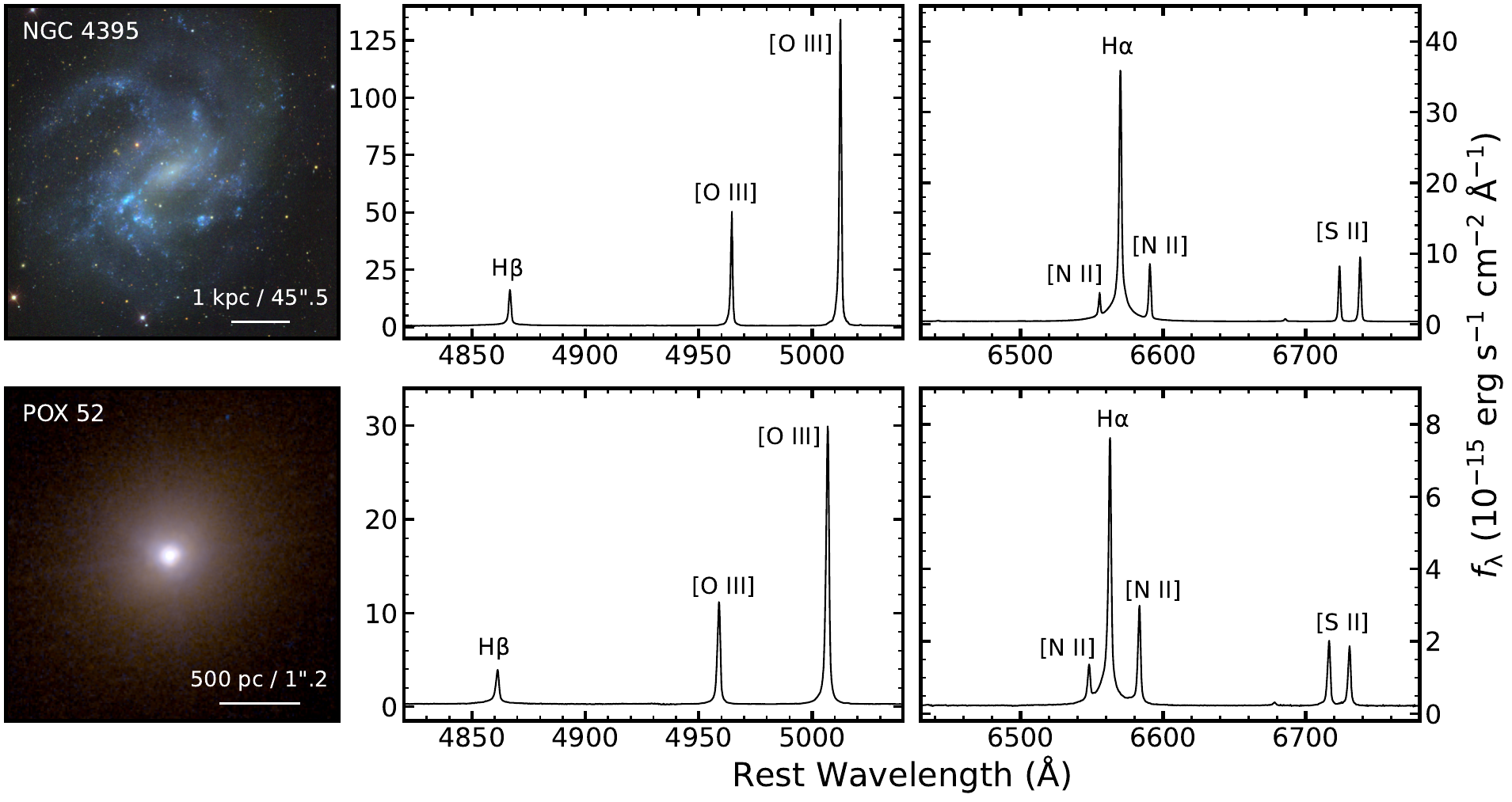}
     \caption{Optical images and spectra of NGC 4395 (top) and POX 52 (bottom). The NGC 4395 image is a composite from SDSS $gri$ bands, while the POX 52 image combines HST ACS/HRC observations with the F814W and F435W filters from \citet{Thornton2008}. The spectra for both targets are obtained from Keck ESI observations as reported in \citet{Barth04}. 
     }
    \label{fig:image_spec}
\end{figure*}

\subsection{IMBH-RM project} 
During the past two decades, minute-level continuum and BLR lags have been successfully detected in NGC 4395, which motivates us to extend this to other IMBH candidates allowing for a sample study. We thus launched the intermediate-mass black hole reverberation mapping (IMBH-RM) project (PI: H. Guo), which includes both continuum and broad-line RM. Further details will be provided in Guo et al. (in preparation), here we give a brief introduction.

We plan to conduct spectroscopic/photometric RM for a selected group of robust IMBH candidates. These candidates exhibit broad Balmer lines and have single-epoch BH masses below $10^6$ M$_{\odot}$, as identified in previous literature \citep[e.g.,][]{Greene04,Greene07,Dong12,Chilingarian18,Liu18}. Our goal is to select $10-20$ most promising IMBH candidates for further RM campaigns, based on multi-line evidence such as optical variability, X-ray/radio detections, and the presence of coronal lines in their spectra. 

Our objectives are as follows: (1) precisely measure continuum and broad-line lags to determine BH masses and confirm their IMBH nature; (2) investigate the variability mechanisms within the BH-accretion disk system, testing the standard thin disk model \citep{Shakura1973}, and exploring possible scenarios such as the X-ray reprocessing scenario \citep{Krolik1991,Cackett2007} or thermal fluctuations \citep{Dexter11,Cai18,Sun20}; (3) assess the applicability of empirical relations established for SMBHs, such as the $M_{\rm BH}-\sigma$ and $M_{\rm BH}-M_{*}$ relations, in the low-mass regime. Given such a sample of IMBHs, it will offer new insights into BH seeding process and the formation history of SMBHs in the early Universe \citep{Greene20}. Among our IMBH candidates, NGC 4395 and POX 52 stand out as the two most reliable cases, supported by multi-band evidence as well as their proximity and brightness, and thus were the first to draw our attention in this study.

\subsubsection{NGC 4395}
NGC 4395 \citep[R.A., Dec. = 186.453583, 33.546864, $\rm D = 4.3 \pm0.4$ Mpc,][]{Thim2004} is a well-known nearby late-type, bulgeless, spiral dwarf galaxy  \citep{Filippenko1989,Filippenko1993} hosting an IMBH \citep{Filippenko2003}, and its optical image and spectrum are shown in Figure \ref{fig:image_spec}. This galaxy hosts the least luminous known Seyfert 1 nucleus, which notably exhibits clear AGN characteristics, such as broad emission lines \citep[$\sigma_{\rm line}$ $\sim$586 km~s$^{-1}$ for \ha,][]{Cho21}, and rapid variability across X-ray, UV/optical, and IR wavelengths \citep{Lira1999, Moran2005, Minezaki06, Edri12, Burke20}. Moreover, continuum and BLR RM signals have been significantly probed \citep{Peterson05, Desroches06, Woo19, Montano22}. Direct measurements of the BH mass through dynamical modeling and RM suggest a mass ranging from $\sim$10$^4$ to a few $\times$ 10$^5$ M$_{\odot}$ \citep{Peterson05, denBrok15, Brum19, Woo19, Cho21}.

\subsubsection{POX 52}
POX 52 (R.A., Dec. = 180.737197, $-$20.934153, $D_{\rm L}$ $\sim$ 93~Mpc) is another nearby dwarf galaxy that hosts an IMBH (Figure \ref{fig:image_spec}). It features a broad \hb\ emission line \citep{Kunth87}, with a width of $\sigma_{\rm line}\sim 323$ \kms\ \citep{Barth04}, which corresponds to a BH mass of a few $\times$ $10^5\ M_{\odot}$. The host stellar velocity dispersion is reported to be $\sim$ 36 $\pm$ 5 \kms\ \citep{Barth04}, independently providing a comparable BH mass according to the $M_{\rm BH}-\sigma$ relation. Clear X-ray and radio variations have been detected in POX 52 \citep{Kawamuro24, Yuan24}. Unlike NGC 4395, POX 52 is a dwarf elliptical galaxy lacking any obvious spiral or disk-like structures, as well as any clumps indicative of star formation. In addition, its accretion rate is close to the Eddington limit, resembling a standard accreting system in classical Seyfert galaxies ($L/L_{\rm Edd}$ $\sim$ 0.5 for POX 52; \cite{Thornton2008}, and $\sim$ 0.05 for NGC 4395; \cite{Moran1999}), offering a unique opportunity to explore a different AGN parameter space.

\subsection{LCOGT Observations for POX 52}
In an effort to probe fast variability, particularly on timescales of hours similar to NGC 4395, we conducted daily-to-minute cadence monitoring of POX 52 using the MuSCAT 4 camera \citep{Narita2020} mounted on the 2-meter Faulkes South Telescope at Siding Spring Observatory in the Las Cumbres Observatory Global Telescope (LCOGT) network. This four-channel camera carry out imaging in $g^\prime$, $r^\prime$, $i^\prime$, and $z_s$ bands ($griz$ hereafter) simultaneously with a field of view (FoV) of 9\farcm 1$\times$9\farcm 1 and a pixel scale of 0\farcs27, providing sufficient sampling to capture minute-level variations. The monitoring was carried out from March 11 to May 21, 2024 UT, with $1-3$ epochs scheduled per day with a single-epoch exposure of 150 s per band. However, some observational requests were not performed due to weather conditions or scheduling competing. This results in 36 epochs were obtained in total for the long-term mode, where the seeing ranges from 1\farcs2 to 3\farcs5, with a median value of 2\farcs0. 

Moreover, to investigate the potential intranight variability, we performed continuous monitoring with a minute-level cadence over a total duration of approximately 8 hours on April 11, 2024 UT (a clear, dark night with longest visible time during a year), similar to \cite{Montano22}. During the intranight monitoring, the seeing ranges from 1\farcs4 to 3\farcs3, with a median value of 2\farcs1. We employed the fast readout mode with a readout time of 6 seconds. Each single-epoch exposure was fixed at 150 seconds per band. In total, we acquired 187 epochs in each band. 

All the raw images are reduced by the BANZAI pipeline\footnote{\url{https://lco.global/documentation/data/BANZAIpipeline/}}provided by LCOGT. Differential photometry was performed on both the target and comparison stars to extract the differencing-image light curves, following the procedure outlined in \citet{Zuo24}. This process included point-spread function (PSF) kernel estimation, image alignment, difference imaging with {\tt HOTPANTS} \citep{Becker2015}, and forced aperture photometry. The average signal-to-noise ratio (SNR) is 147, 250, 143, and 99 for $griz$ bands, respectively.

\begin{figure*}[ht!]
    \centering
    \includegraphics[width=0.9\linewidth]{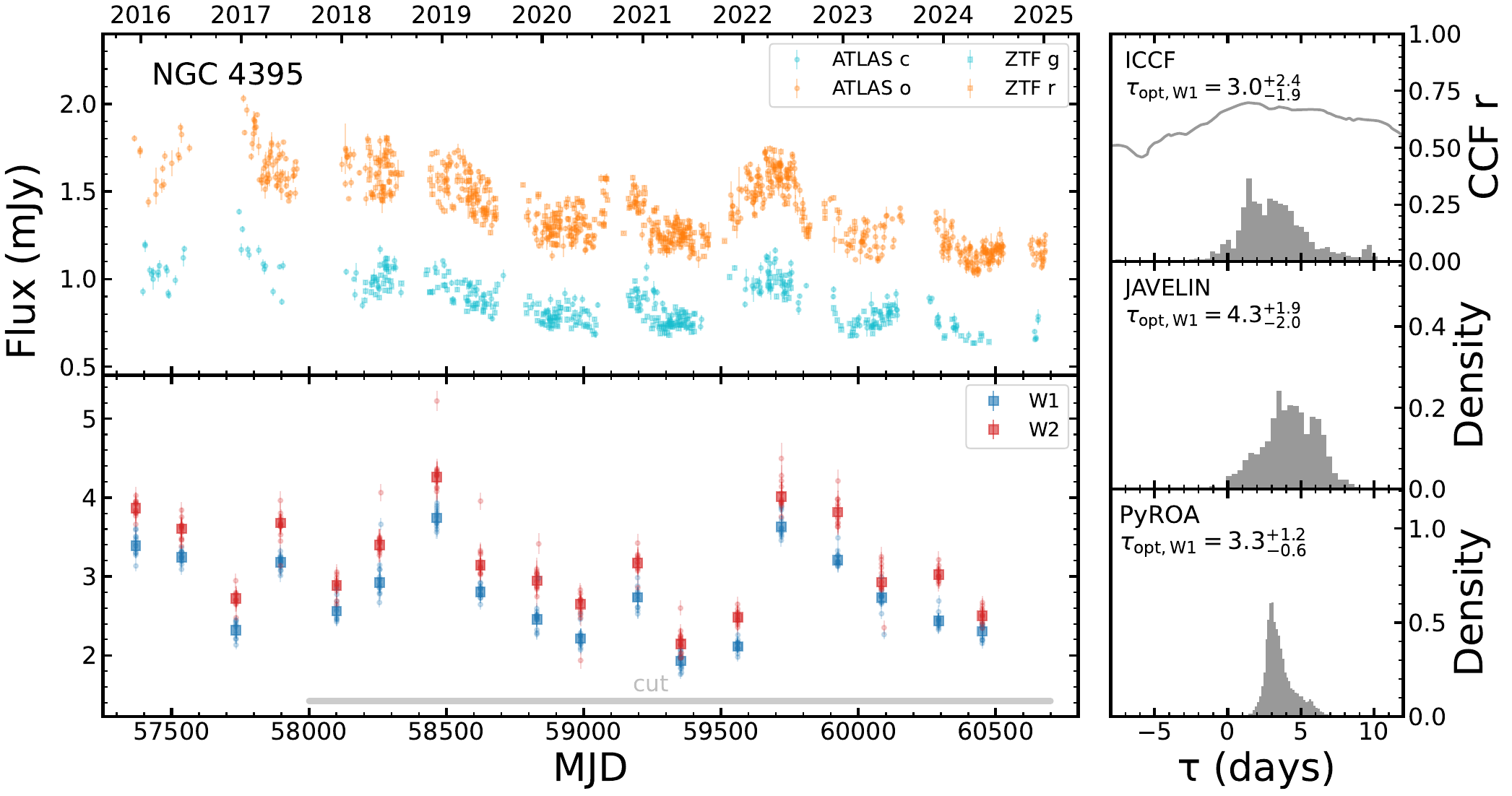}
    \bigskip
    \includegraphics[width=0.9\linewidth]{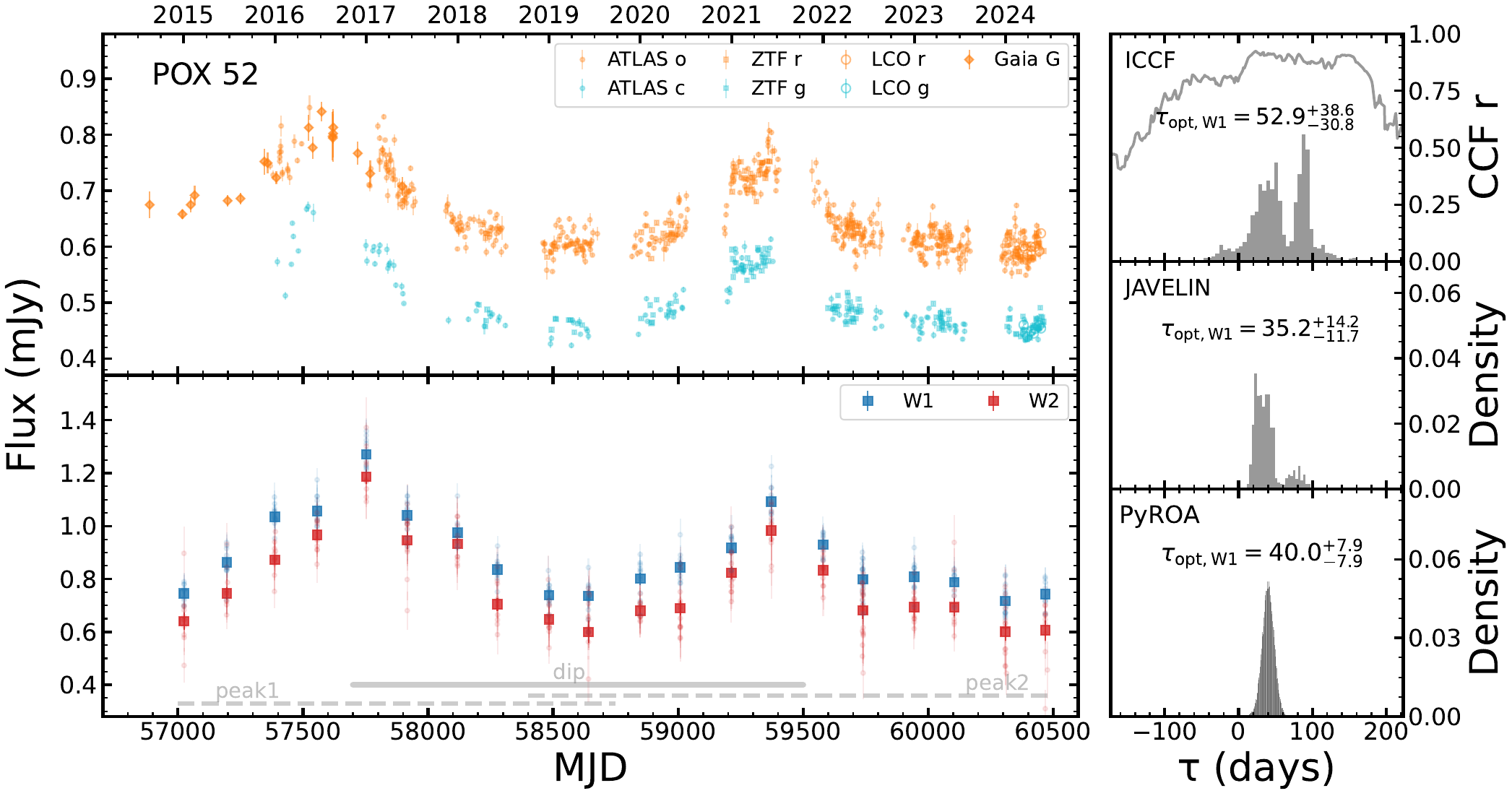}
    \caption{Light curves and time lag measurements of NGC 4395 and POX 52. The left panels display the combined optical light curves from multiple telescopes alongside the original (dots) and binned (squares) MIR WISE light curves for each target. The CCF curves, and log posterior distributions of ICCF, JAVELIN, and PyROA in the fiducial segments are displayed in the right panels. Note that the false peak of JAVELIN in POX 52 was removed, see Appendix \ref{app:alias}.}
    \label{fig:opt_lc_2band}
\end{figure*}

\subsection{Optical Light Curves}
The Asteroid Terrestrial-impact Last Alert System \citep[ATLAS,][]{Tonry18} consists of four 0.5-meter telescopes (two in Hawaii, one in Chile, and one in South Africa). Since June 2015, ATLAS has been automatically scanning the entire sky with an average cadence of one day in the $c$ ($\lambda_{\rm ref} = 5288$ \AA; $W_{\rm eff} = 2145$ \AA) and $o$ ($\lambda_{\rm ref} = 6750$ \AA; $W_{\rm eff} = 2368$ \AA) bands, aiming to detect hazardous near-Earth asteroids. PSF-forced photometry of NGC 4395 and POX 52 on the reduced images is obtained from the ATLAS photometry servers\footnote{\url{https://fallingstar-data.com/forcedphot/}}.

Zwicky Transient Facility (ZTF), an automated time-domain survey, utilizes the 1.2-meter (48-inch) Schmidt telescope at Palomar Observatory, which has an exceptionally wide field of view ($\sim$47 deg$^2$) \citep{Bellm19}. Its primary goal is to monitor the entire variable sky, covering a broad range of time-domain science. Since March 2018, ZTF has been surveying the entire sky visible from Palomar (Dec $>-30^{\circ}$) with a cadence of $\sim$ 2 days in the $gri$ bands. Light curves of NGC 4395 and POX 52 are generated from calibrated single-exposure PSF-fit photometry and can be accessed via the NASA/IPAC Infrared Science Archive\footnote{\url{https://irsa.ipac.caltech.edu/docs/program_interface/ztf_lightcurve_api.html}}. For ATLAS and ZTF, photometric data with seeing worse than 4\arcsec\ or SNR below 10 are abandoned in further analysis. 

Additionally, the Gaia G-band ($\lambda_{\rm ref} = 6218$ \AA; $W_{\rm eff} = 4053$ \AA) light curve of POX 52 was obtained from the Gaia Photometric Science Alerts Website\footnote{\url{https://gea.esac.esa.int/archive/}} \citep{Hodgkin21}, where the photometry is performed using Line Spread Function (LSF) fitting. The Gaia data used in this study spans from August 2014 to May 2017, with an average cadence of approximately two months.

To extend the baseline and enhance the cadence of optical light curves, new blue and red band light curves are constructed by combining data from multiple surveys with similar wavelengths. The red band compiles ATLAS $o$ (6750 \AA), ZTF $r$ (6417 \AA), and Gaia $G$ (6218 \AA), while the blue band combines ATLAS $c$ (5288 \AA) and ZTF $g$ (4783 \AA). Although the Gaia $G$ band spans a broad wavelength range, it is combined with the ATLAS $o$ band and ZTF $r$ band to leverage their superior cadence and sensitivity. The resulting red-band light curve is then used to analyze the optical variability. For POX 52, additional calibration was required due to flux mismatches across surveys. ZTF and Gaia fluxes were scaled and shifted to match contemporaneous ATLAS fluxes, with scaling and offset factors of 0.81/0.15 mJy for ZTF $g$, 0.84/0.17 mJy for ZTF $r$, and 0.96/0.43 mJy for Gaia $G$. For NGC 4395, the flux levels and variability amplitudes in the ZTF and ATLAS light curves are consistent, and no further calibration is necessary.

\subsection{MIR Light Curves}
WISE is a space telescope that began its comprehensive all-sky IR survey in December 2009. After completing its primary mission, WISE was reactivated in 2013 as the NEOWISE mission, focusing on monitoring near-Earth objects. During this phase, it continued periodic sky scans, typically every six months, producing deep and continuous MIR monitoring, until its recent retirement in 2024. These observations were conducted in the W1 (3.4 $\mu$m) and W2 (4.6 $\mu$m) bands. All PSF-fit photometries of NGC 4395 and POX 52 are available for download from the NASA/IPAC Infrared Science Archive\footnote{\url{https://irsa.ipac.caltech.edu}}. We abandoned those photometries flagged as poor image quality (\texttt{$\rm qi\_ fact=0$}), moonlight scattering (\texttt{$\rm moon\_masked=0$}), or contamination of artifacts (\texttt{$\rm cc\_flags \neq 0000$}). Each half-yearly visit spans one to several days, during which multiple epochs are combined to enhance photometric accuracy, as most measurements exhibit negligible variability. Four notable exceptions in NGC 4395 exhibit potential variability at the $\gtrsim$0.1 mag level and are discussed in Appendix~\ref{app:MIR_daily}.

\subsection{Lag measurements} \label{sec:measure_lag}

\textbf{ICCF} (Interpolated cross-correlation function) is a commonly used method to calculate the time lag by interpolating the light curves and identifying a time delay at which their correlation is maximized \citep{Peterson98}. In this work, the ICCF measurements are performed using a Python-based improved interpolated cross-correlation function\footnote{\url{https://github.com/legolason/PyIICCF}} \citep[PyI$^2$CCF; see][for details]{Guo21,U22}. The lag search range is set to 10$\times$ the predicted MIR lag to mitigate the potential alias effect, with an interpolation grid spacing of 0.1 day. Specifically, the \ha\ lag (similar to that of \hb) of NGC 4395 is less than 3 hours \citep{Cho21}, corresponding to an MIR lag at the W1 band of approximately 30 hours assuming the torus size at W1 is $\sim$ 10$\times$ BLR size \citep{Chen23}. Thus, we adopt an MIR search range around 10 times larger, i.e., $\pm$15 days. Regarding POX 52, given its AGN continuum luminosity of $\sim$10$^{42}$ erg s$^{-1}$ at 5100\AA, the predicted BLR lag is around 30 days based on the $R_{\rm BLR}-L$ relation \citep{Wang24}. Therefore, we set an MIR search range of $\pm$300 days. The centroid lag is calculated by weighting all points within 80\% of the maximum value of the CCF, which provides a more representative estimate of the response radius of the variability signal \citep[e.g.,][]{Gaskell86}. A Monte Carlo method is employed through both flux randomization (FR) and random subset sampling (RSS) with 50,000 iterations \citep{White94,Peterson98}. The median of the posterior distribution is adopted as the measured time lag, with the 1$\sigma$ range representing the uncertainty.

\textbf{JAVELIN} is another widely used tool for measuring time lags \citep{Zu11}. It models the optical light curve using a damped random walk (DRW) model, which is generally a good representation of AGN variability \citep{Kelly09,MacLeod2010}, albeit with deviations on very short (minutes) and long (decades) timescales \citep{Mushotzky11,Guo17,Su24a}. Unlike the simple linear interpolation of light curves in ICCF, JAVELIN assumes that the quasar's statistical variability follows the DRW model and interpolates the light curves accordingly. This approach could provide more reliable time lag measurements with reduced influence from the seasonal gaps in the light curves \citep{Li19,Yu20}. The method employs a Markov Chain Monte Carlo (MCMC) approach to model the driving and responding light curves, where the responding light curve is treated as a shifted, scaled, and smoothed version of the driving light curve. We caution that this assumption could be oversimplified \citep{Goad16, Homayouni23} and the transfer function of dust reprocessing may be more complex, owing to the poorly understood structure and dynamics of the torus \citep{Lyu2022}. The lag search range is set to match that in the ICCF method, with 50,000 MCMC chains used to generate the lag posterior.

\textbf{PyROA} adopts a running optimal average (ROA) to model the variability of AGN light curves \citep{Donnan2021}. It adopts an empirical approach where the continuum and the shifted/scaled MIR light curves are jointly fitted using a Gaussian-windowed ROA model. The time lag, the Gaussian window width, and other model parameters are simultaneously constrained through Bayesian inference, balancing model complexity with goodness of fit. In our implementation, we adopted 50,000 MCMC samples in the same search range as the ICCF and visually checked the MCMC chain to be convergent. The ICCF-measured lag was used as an initial guess to assist convergence. Extra variance for the uncertainty is added as a parameter to mitigate the influence of significant outliers. PyROA has recently demonstrated accurate lag measurements in the SDSS-RM project \citep{Shen24}, with the advantage of being model-independent. However, PyROA results are sensitive to initial parameter values, and the application of Gaussian smoothing can sometimes lead to overfitting, particularly in sparsely sampled light curves \citep{Wang24}. Therefore, in this work, PyROA is used to validate the ICCF and JAVELIN lag measurements, serving as a consistency check rather than a primary estimator.

\section{Results}\label{sec:results}

\begin{figure*}[t!]
    \centering
    \includegraphics[width=1\linewidth]{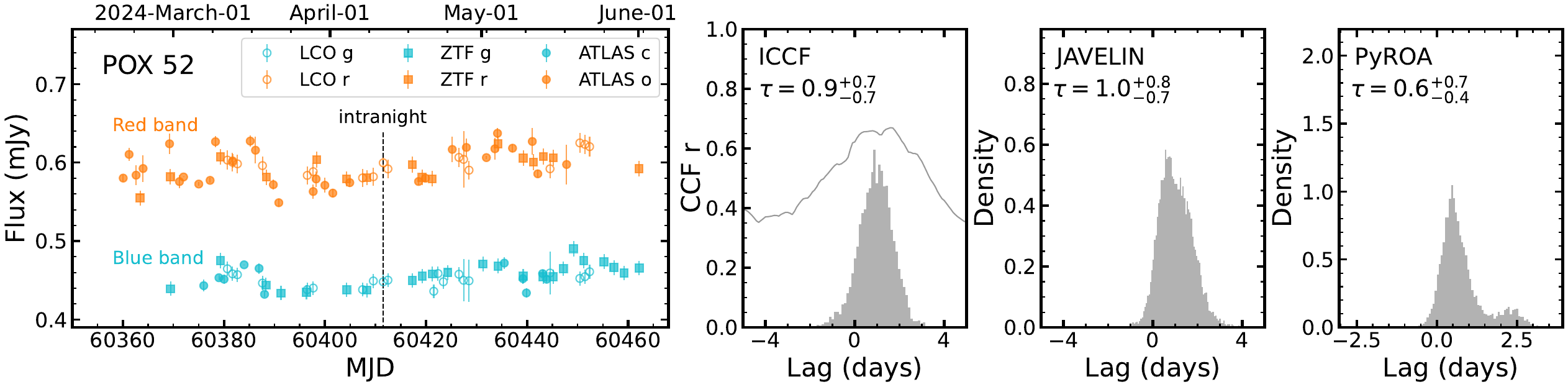}
    \caption{Left panel: the zoom-in light curve of POX 52 during the LCOGT monitoring. The dotted line indicates the time of intranight monitoring. Right panels: ICCF, JAVELIN, and PyROA results. }
    \label{fig:LC_monitoring}
\end{figure*}

\begin{figure}
    \centering
    \includegraphics[width=1\linewidth]{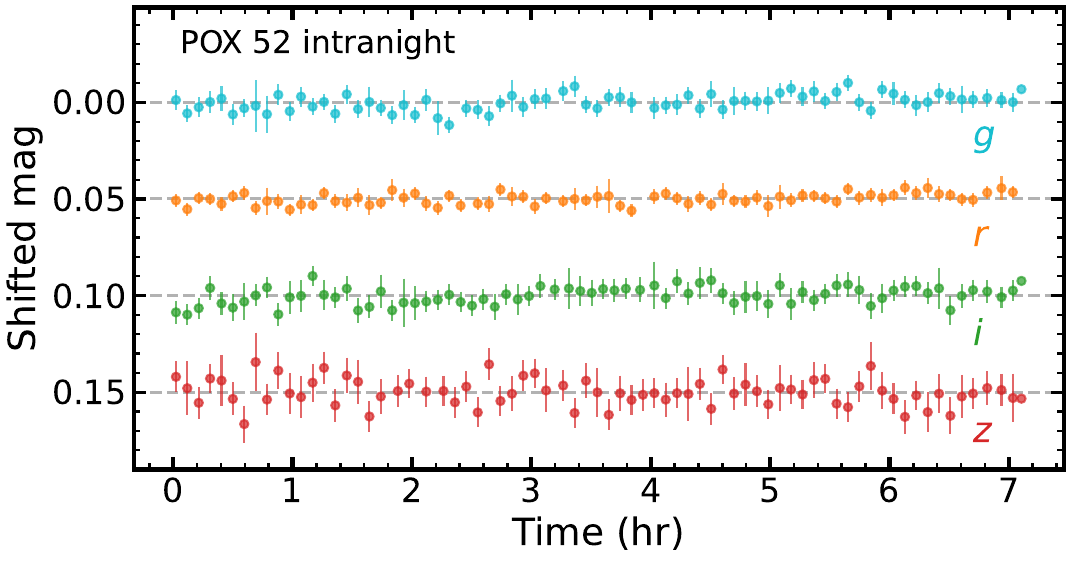}
    \caption{Multi-band optical monitoring of POX 52 on April 11 2024. MuSCAT 4 light curves in the $griz$ bands. No significant intranight variability was detected.}
    \label{fig:intranight}
\end{figure}

\begin{deluxetable*}{lccccccccc}
\tablecaption{Lag results in the observed frame for NGC 4395 and POX 52. }
\tablehead{ \colhead{Name} & \colhead{MJD range} & \colhead{$r_{{\rm max, W1}}$}  & \colhead{$r_{{\rm max, W2}}$} & \colhead{$\tau_{\rm ICCF,W1}$}  & \colhead{$\tau_{\rm ICCF, W2}$}  & \colhead{$\tau_{\rm JAV,W1}$}  & \colhead{$\tau_{\rm JAV, W2}$} & \colhead{$\tau_{\rm ROA, W1}$} & \colhead{$\tau_{\rm ROA, W2}$} \\
 & & & & (day) & (day) & (day) & (day) & (day) & (day) }
\startdata
NGC 4395 all & 57400$-$60700 & 0.58    & 0.58    & $2.4^{+1.3}_{-1.5}$       & $2.5^{+1.7}_{-1.5}$       & $4.3^{+1.7}_{-1.6}$     &  $4.6^{+2.9}_{-3.6}$ & $5.2^{+1.1}_{-1.2}$ & $5.0^{+2.6}_{-1.4}$ \\
\textbf{NGC 4395 cut} & 58000$-$60700 & 0.70    & 0.69    & $\mathbf{3.0^{+2.4}_{-1.9}}$      & $3.3^{+3.1}_{-2.2}$       & $4.3^{+1.9}_{-2.0}$     & $4.2^{+3.2}_{-2.7}$   & $3.3^{+1.3}_{-0.6}$  & $2.7^{+0.5}_{-0.8}$ \\
\hline 
POX 52 all   & 57000$-$60500 & 0.92    & 0.92    & $86.5 ^{+45.7} _{-43.8}$  & $128.1 ^{+42.7} _{-62.0}$ & $88.7^{+13.6}_{-21.8}$ & $97.7^{+15.9}_{-27.4}$ & $36.5^{+8.8}_{-9.3}$  & $24.5^{+19.6}_{-27.5}$\\
POX 52 peak1 & 57000$-$58750 & 0.93    & 0.94    & $138.4 ^{+32.1} _{-37.1}$ & $159.7 ^{+60.1} _{-40.6}$ & $95.8^{+17.9}_{-19.8}$ & $114.6^{+17.9}_{-28.8}$  & $166.2^{+10.4}_{-10.2}$ & $163.7^{+16.1}_{-17.2}$        \\
{\bf POX 52 dip}   & 57700$-$59500 & 0.92    & 0.92    & $52.9^{+38.6}_{-30.8}$    & $84.2^{+34.1}_{-46.6}$    & $\mathbf{35.2^{+14.2}_{-11.7}}$  & $35.3^{+33.9}_{-11.9}$ & $39.7^{+8.0}_{-8.4}$ & $35.9^{+14.2}_{-18.4}$\\
POX 52 peak2 & 58400$-$60500 & 0.94    & 0.93    & $32.6^{+49.9}_{-70.6}$    & $67.9^{+78.8}_{-80.3}$    & $6.3^{+9.3}_{-9.1}$  & $16.0^{+32.2}_{-22.3}$ & $12.6^{+11.7}_{-13.3}$ & $9.7^{+27.5}_{-38.6 }$ 
\enddata
\tablecomments{The lags are estimated using the ICCF, and JAVELIN, and PyROA methods applied to different light-curve segments between the red optical band and the WISE W1 \& W2 bands. Note that false peak has been removed in the JAVELIN lag estimation of POX 52 (see Appendix \ref{app:alias}). The ICCF lags for NGC 4395 (cut) and the JAVELIN lags for POX 52 (dip), shown in bold, are considered as the fiducial ones ; further details are provided in the text.}
\end{deluxetable*}\label{tab:lag}

\subsection{Lags between optical and MIR }
The dusty torus is suggested to have a stratified structure, and its thermal MIR emission is driven by the UV/optical continuum \citep{Netzer15}. As shown in Figure \ref{fig:opt_lc_2band}, the MIR variations clearly respond to the optical variations, albeit with the sparse WISE cadence. Both the optical and MIR light curves exhibit significant long-term and typical-AGN variations near a decade baseline. The variability amplitude ratios between maximum and minimum of NGC 4395 and POX 52 are $\sim$ 2 (1.8) and $\sim$ 1.4 (1.7) in the optical (MIR) bands, respectively. The even stronger MIR variation relative to optical in POX~52 could be due to the high covering factor, large inclination angle of the optical emission (more edge on), or dust extinction in optical \citep{Stalevski16}. On the other hand, the optical light curves of NGC 4395 exhibit more pronounced rapid variations than those of POX 52, likely due to its lower luminosity and BH mass, which correspond to shorter variability timescales \citep{Burke21}.

We performed the lag measurements for the overall light curves and different segments with significant variable features using ICCF, JAVELIN, and PyROA, all the results are listed in Table \ref{tab:lag}. 

NGC 4395 generally exhibits a moderate correlation between optical and MIR data, with cross-correlation coefficients exceeding 0.55. The overall lag correlation is relatively lower due to an apparent anti-correlated trend observed before 2018. However, this trend is likely artificial, resulting from sparse data sampling and large seasonal gap during the early stages of the ATLAS survey. Remarkably, a small MIR lag of $\sim$ 3 days in NGC~4395 is detected, given the significant variation and high-cadence (2-day) optical light curves, demonstrating the feasibility of detecting the short lags with a combination of high- and low-cadence light curves. We adopt the ICCF lag of $3.0^{+2.4}_{-1.9}$ days for NGC 4395 (cut) in Table \ref{tab:lag} as the fiducial value, given its model-independence and stronger correlation than the overall result. It is consistent with the JAVELIN and PyROA lags within 1$\sigma$ uncertainty, and the Gaussian-window width of $\sim$1.7 days also supports a compact dust torus in NGC 4395.

POX 52 exhibits two major symmetric peaks, indicative of typical AGN variability rather than transient events. The time lag measurements from all methods reveal considerably longer MIR lags for POX 52 compared to NGC 4395, consistent with its higher luminosity. The lag measurements vary across different light-curve segments, particularly between the first and second peaks. We attribute this variation primarily to the alias effect, exacerbated by the large seasonal gaps ranging from 120 to 160 days in the optical data and a half-year gap in the MIR data, rather than the intrinsic lag variation \citep{Su24}. As a result, the measured lags in each segment differ across methods due to the influence of the aliasing peak, as detailed in Appendix \ref{app:alias}. Figure \ref{fig:POX52_post} shows that most lag results are heavily affected by this alias-induced false peak. Fortunately, in the dip segment, the false peak is clearly separated, and the alias-reduced lag still provides a convincing lag, with all three methods yielding lags around 40 days\footnote{The ICCF result (dip) is still slightly affected by a possible alias peak near 95 days; removing it brings the result in line with JAVELIN and PyROA.}. We adopt the JAVELIN lag of $35.2^{+14.2}_{-11.7}$ days (dip) as the fiducial value, as it more effectively reduces the impact of aliasing and provides a more significant lag detection than other segments. This result is also confirmed by PyROA with a retrieved time lag of $\sim$40 days and a Gaussian-window width of around 30 days. Attempts to narrow the lag search window did not significantly reduce aliasing, and no notable lag difference between W1 and W2 bands was found in either object, given the large uncertainties.

\subsection{Optical Intranight Variability of POX 52} \label{sec:monitoring}
As the intranight variability of NGC 4395 has been consistently confirmed in previous studies, we also performed optical monitoring of POX 52 through LCOGT, incorporating both long-term and short-term (intranight) modes, with the aim of detecting potential fast variations. The optical continuum lag of NGC 4395 is $\tau_{\rm gz} \sim 20$ minutes \citep{Montano22}, and considering the $\sim$200$\times$ higher continuum luminosity in POX~52, the inter-band lags in the optical are predicted to be $\sim$ 0.3 day, given $\tau_{\rm gz} \propto L^{0.56}$ \citep{Montano22}.

Figure~\ref{fig:LC_monitoring} presents observed light curves of POX 52 over three months. These dual-band light curves reveal a strong correlation, with $r_{\rm max} \sim 0.7$ and a variation of $\sim$10\%. Notably, mild lags between the blue and red bands are detected by all the ICCF, JAVELIN, and PyROA methods, with $\tau_{\rm ICCF}$ = 0.9$^{+0.7}_{-0.7}$ days, $\tau_{\rm JAV}$ = 1.0$^{+0.8}_{-0.7}$ days, and $\tau_{\rm ROA} = 0.6^{+0.7}_{-0.4}$ days in the observed frame, generally consistent with previous prediction. 

No robust intranight variations were observed, as shown in Figure \ref{fig:intranight}. We attribute this to the limited monitoring duration, which was insufficient to capture significant variability. In other words, POX 52 may intrinsically have a lower probability of exhibiting extremely fast variations comparing with NGC 4395, as the characteristic variability timescale is expected to scale with the BH mass \citep{Burke21}, reducing the likelihood of detecting intranight variations (for further discussion, see \S \ref{sec:micro_var}). 

\subsection{Caveats of lag measurements}

In our campaign, we emphasize that the cadence of the WISE light curves is significantly lower than that of the optical light curves. We assess the time lag recovery using mocked light curves for NGC~4395 and POX~52, with details provided in Appendix~\ref{app:mock}. For both sources, the input time lag is well recovered within \(1\sigma\), adopting a deliberately more stringent definition of successful recovery compared to previous studies \citep[e.g., within \(3\sigma\),][]{Li19}. The recovery rates for both NGC~4395 and POX~52 are about 50\%. In addition, the dip segment in POX~52 exhibits the smallest scatter and appears least affected by seasonal gaps, thus adopted as the fiducial lag results.

Moreover, we follow the traditional bidirectional interpolation used in the ICCF method, which tends to overestimate the lag uncertainty and substantially reduces the cross-correlation coefficient. In ICCF measurements, the random subset sampling method for error estimation discards about one-third of the data points during the Monte Carlo process \citep{Peterson98}. This is particularly problematic for the sparse WISE light curves, leading to a substantial increase in uncertainty. Additionally, JAVELIN is a model-dependent method that assumes the reprocessed light curve is a scaled, smoothed, and shifted version of the driving light curve. However, this is not a good assumption, given some significant MIR outbursts are found without the optical counterpart \citep{Jiang2021}, possibly due to the anisotropic illumination or self-shielding within the torus. Finally, the lag uncertainties from JAVELIN and PyROA are known to be systematically smaller than those derived from the ICCF method \citep{Li19, Yu20, Wang24}.

\begin{figure*}[ht!]
    \centering
    \includegraphics[width=0.95\textwidth]{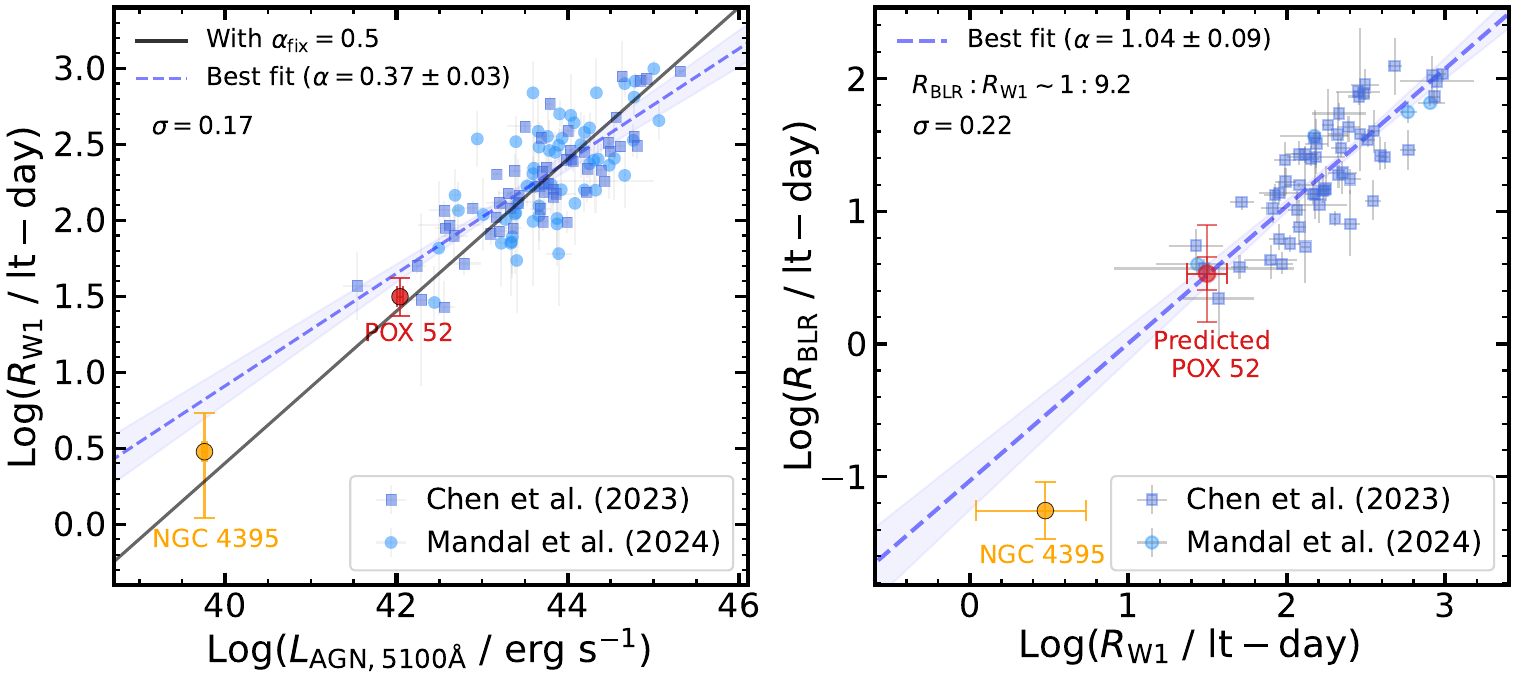}
    \caption{Left panel: correlation between the dust size inferred from $W1$ band and the AGN 5100 \AA\ luminosity. The solid line represents the fit with $\alpha$ fixed at 0.5, while the blue dashed line shows the best fit, with the shaded region indicating the 1 $\sigma$ uncertainty, with an additional scatter of 0.17 dex. Right panel: correlation between the dust size and the BLR size. The solid line represents the fitted linear relation. The dashed line is the best-fit and the shaded region indicate the 1 $\sigma$ uncertainty. The red point represents the predicted BLR lag of POX 52 on the linear relation, with the 0.15 measurement error (inner bar) plus 0.22 systematic error (outer bar). }
    \label{fig:RL}
\end{figure*}

\section{Discussion} \label{sec:diss}
\subsection{Extending R$_{\rm torus}$--L$_{\rm AGN}$ relation to IMBH regime}
According to the radiation equilibrium of dust \citep{Barvainis87}, the dusty torus size is determined by the sublimation radius of grains such as graphite and silicate, which expects a size--luminosity relation of $R_{\rm torus}\propto L_{\rm AGN}^{0.5}$. However, a shallower relation with an index of $\sim$0.3 to 0.4 is found in both dust RM studies \citep{Yang20, Chen23, Mandal24} and NIR interferometric measurements \citep{Gravity2024}.
Several possibilities have been proposed to explain this deviation: First, SMBHs accreting at super-Eddington rates may exhibit shorter lags due to the self-occultation effect in the slim disk model \citep{Kawaguchi11}. Second, in addition to the cosmic time dilation effect, the observed lag likely corresponds to the lag at different rest-frame wavelengths, with longer wavelengths typically associated with longer lags due to the more extended emitting regions. Third, due to the limited baseline, the lags detected in objects tend to be smaller \citep{Yang20}. Finally, the AGN continuum luminosity at 5100\AA\ may not accurately represent the bolometric luminosity, while applying a non-linear bolometric correction factor could retrieve the canonical $R_{\rm torus}\propto L_{\rm AGN}^{0.5}$ relation \citep{Gravity2024}. 

To mitigate potential biases caused by the Eddington ratio and redshift, we compiled a subsample from existing MIR RM literature with Eddington ratio estimates \citep{Chen23, Mandal24}. This subsample is restricted to sources with redshift $z < 0.2$ and Eddington ratio $\dot{m} < 0.5$, resulting in 107 objects, where the mean values are adopted for seven duplicated objects. Following \cite{Mandal24}, the redshift correction $(1+z)^{-0.38}$ is applied to the MIR W1 lag, accounting for the combined effect of cosmological time dilation and the wavelength dependence of the dust time lag. By including NGC 4395 and POX 52, we extend the analysis into a low-mass regime, which exhibits luminosities two orders of magnitude lower than those of SMBHs, as shown in the left panel of Figure \ref{fig:RL}. The AGN luminosity of NGC 4395 at 5100 \AA\ is measured as $\lambda L_{\rm \lambda,5100} = (5.75 \pm 0.4)\times 10^{39}\rm \ erg\ s^{-1}$, based on the the rescaled mean spectrum with host correction \citep{Cho20}. For POX 52, we adopt the $\lambda L_{\rm \lambda,5100} = 1.1\times 10^{42}\rm \ erg\ s^{-1}$, determined from a joint decomposition using spectral energy distribution (SED) and image with reported uncertainty of a few percentages \citep{Kawamuro24}.

In order to parameterize the $R_{\mathrm{torus}}-L_{\rm AGN}$ relation, we apply a linear MCMC fitting for the combined samples using the python package {\tt emcee} \citep{Foreman-Mackey13}, by fitting the function with a free and fixed slope ($\alpha = 0.5$):
\begin{equation}
    \log({R}_{\mathrm{torus}}/\textrm{1 lt-day}) = \alpha\ \log \left (\frac{{L}_{\mathrm {AGN}, 5100}} {10^{43}\ \textrm{erg\ s}^{-1}} \right ) + K.
\end{equation}
The best fit gives $K = 1.75^{+0.01}_{-0.01}$ with $\alpha$ fixed to 0.5 and $\alpha = 0.37^{+0.03}_{-0.03}$, $K = 2.03^{+0.03}_{-0.03}$ for free fit. The latter confirms a shallower slope consistent with previous results \citep{Yang20,Chen23,Mandal24}, yet with only one extra weight-limited data point at the low-luminosity end. This does not necessarily indicate that the slope of this relation deviates from 0.5, as NGC~4395 and POX~52 are located between these two correlations. Furthermore, if previous samples properly account for the potential contributions of the accretion disk to the MIR emission, the relation could more closely reflect the predicted slope \citep{Lyu19}. The inclusion of more low-luminosity IMBHs will further reinforce this relation.

\subsection{MIR lag-inferred BH mass of POX 52}
The right panel of Figure~\ref{fig:RL} exhibits the correlation between the BLR size and the torus size of 56 sources from \citet{Chen23} and 4 non-redundant objects from \citet{Mandal24} with sub-Eddington and low-redshift including NGC 4395. The best-fit slope is $1.04 \pm 0.09$, consistent with the linear relation reported by \cite{Chen23}, which suggests that the dust torus is systematically 9.2 times larger than the BLR region, with an intrinsic scatter of 0.22 dex. However, NGC~4395 lies below this relation, probably indicating that our MIR lag measurements are slightly overestimated due to cadence limitations or potential evolution in AGN structure as a function of BH mass, which is beyond the scope of this paper. Given the observed $W1$ time lag of $34.5^{+13.9}_{-11.4}$ days in POX 52, the inferred BLR size is $3.7^{+4.4}_{-1.9}$ lt-day considering the uncertainties in time lag measurement and intrinsic scatter of the $R_{\rm BLR}-R_{\rm torus}$ relation. Then the BH mass can be estimated by:  

\begin{equation}
    M_{\rm BH} = f \frac{\sigma_{\rm line}^{2} R_{\rm BLR}}{G}
\end{equation}
where $G$ is the gravitational constant, $f$ is the virial factor, with the adopted value of 4.47 from \cite{Woo15}, and $\sigma_{\rm line}$ is the velocity dispersion of the broad emission line, represented by the H$\beta$ broad component velocity dispersion of 323 $\pm$ 13 km s$^{-1}$ in the single spectrum from \cite{Barth04}. The BH mass is estimated to be $10^{5.5}\ M_{\odot}$ with $\pm$0.15 dex (statistical error) and $\pm$0.22 dex (systematic error), leading to a total uncertainty of 0.37 dex, and thus generally confining the BH mass of POX 52 to $<10^6$ M$_{\odot}$.  

However, the BH mass estimates derived from MIR lags are quite uncertain. Firstly, the MIR lags from light curves and line widths from spectroscopic observations are not obtained simultaneously, and the continuum luminosity and line widths could vary significantly in decades. In addition, since there is only one data point NGC 4395, yet with large uncertainty, in the low-luminosity end of the $R_{\rm torus}-R_{\rm BLR}$ relation, it is difficult to determine the precise slope of the relation, hence the lag prediction of POX 52. For example, recent works suggest the $R_{\rm torus}-R_{\rm BLR}$ relation could be slightly steeper than the linear relation, with $\alpha$ of 1.2 to 1.4 \citep{Gravity2023, Mandal24}, resulting in a smaller BH mass of POX 52. A further caveat is that both the BLR and MIR lags use the optical continuum as a reference, but recent CRM studies have revealed non-negligible inter-band lags, primarily originating from the diffuse continuum of the BLR \citep{Korista19,Guo22}. Consequently, using different reference points may introduce additional bias in the lag measurements.

\begin{figure}
    \centering
    \hspace{-1cm}
    \includegraphics[width=0.5\textwidth]{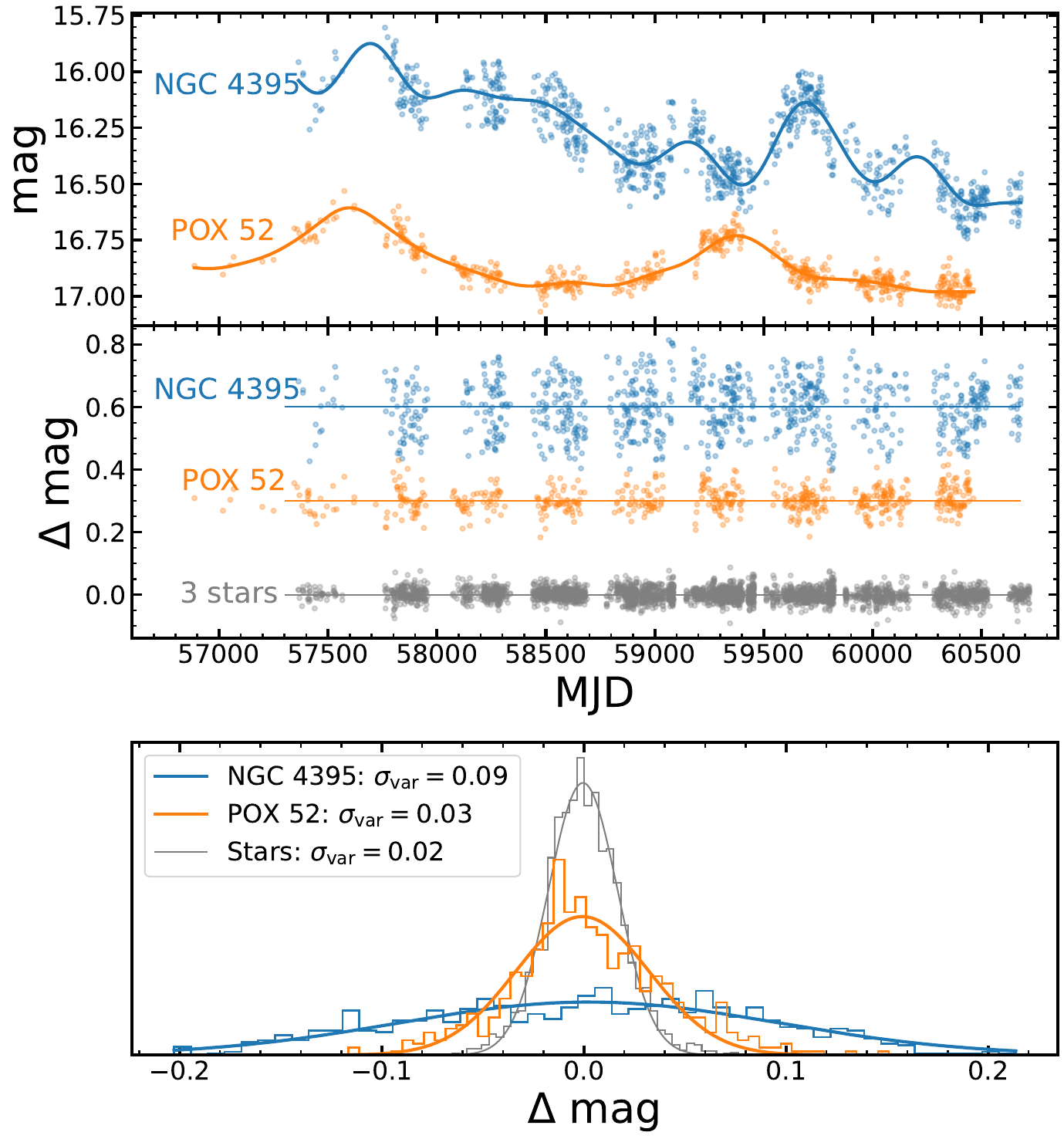}
    \caption{Light curve detrending and evaluation of fast variability in NGC 4395 and POX 52. Top panel: The light curves are detrended using a low-pass filter with a carbox of one year. The detrended residuals for two targets and reference stars are shown. Bottom panel: The distributions of the residuals are fitted with Gaussian profiles. The 1$\sigma$ width of the stars represents the intrinsic scatter, while any additional broadening indicates the potential real variability from two IMBHs. }
    \label{fig:detrend}
\end{figure}

\subsection{Fast variability} \label{sec:micro_var}

Several pioneering studies have explored the rapid variability of IMBHs on timescales ranging from hours to days, typically using relatively sparse sampling at 1–2 hour intervals \citep{Martinez-Palomera20, Shin22}. \citet{Gopal-Krishna23} conducted a systematic search for intranight variability in 12 IMBHs, carrying out 36 monitoring sessions, each lasting more than three hours with minute-level cadence. Their study revealed a detection rate of approximately 20\%, with observed variations in the range of 0.05 to 0.2 mag.

To evaluate the detection rate of the fast variability, we detrended the red band light curves in both NGC 4395 and POX 52 using a low-pass filter with a width of 1 year to remove the long-term variation. Figure \ref{fig:detrend} exhibits the long-term trend and the detrended residual light curves, including three reference stars with comparable brightness ($\sim$ $14 - 17$ $r$ band mag) for comparison in the top panels. The gray histogram in the bottom panel represents the intrinsic scatter of three selected invariable stars obtained through PSF photometry\footnote{PSF photometry is generally suitable for POX 52 and NGC 4395, as POX 52 appears point-like in ZTF and ATLAS, and PSF photometry is applied only to the nucleus of NGC 4395, minimizing scatter. We extensively tested difference image photometry, but despite efforts, its scatter remained larger than that of PSF photometry, especially for ATLAS, leading us to adopt PSF photometry for more accurate lag measurements.}. 


Both NGC 4395 and POX 52 clearly exhibit larger scatter than the reference stars, indicating genuine short-term variability. To isolate the intrinsic fast variability, we subtract the systemic scatter using the stars: $\sigma_{\rm intr} = \sqrt{\sigma^{2}_{\rm var} - \sigma^{2}_{*}}$, where $\sigma_{\rm var}$ and $\sigma_{\rm *}$ represent the 1$\sigma$ scatter of the sources and stars, respectively. The intrinsic scatter of NGC 4395 is found to be 0.088 mag, implying a 68\% chance of observing a variation larger than 0.088 mag over the past decade, which is consistent with previous observations \citep{Woo19,Montano22}. It is important to note that the light curves are binned in 1-day intervals and with a typical 2-day cadence, which means that faster variability signals are smoothed out. As a result, this estimate represents a lower limit, and the actual fast variability could be more frequent and pronounced. For POX 52, the intrinsic scatter is 0.022 mag, indicating much weaker fast variability during the observations.

To further estimate the typical timescale for fast variability, we calculate the average period required for a 0.1 mag variation. Starting from the initial point of the light curve, we identify subsequent points where the difference from the preceding maximum or minimum exceeds 0.1 mag. We then proceed through the light curve in successive windows, counting the occurrences of such variability and calculating their timescale. The median value of the timescale distribution is adopted as the estimated timescale, with 15.9\% and 84.1\% of the distribution representing the uncertainty. This process yields 309 periods for NGC 4395 and 72 periods for POX 52 across the entire light curve. The derived timescale for 0.1 mag variability is $3.9^{+5.1}_{-2.3}$ days for NGC 4395 and $11.8^{+20.1}_{-7.3}$ days for POX 52, suggesting that the duty cycle of fast variability in POX 52 is three times longer than in NGC 4395, which could be a primarily reason of non-detection of intranight variation in \S \ref{sec:monitoring}. Note that the light curve has a 2-day cadence, and if intrinsic fast variability occurs on timescales shorter than the observational sampling, the average timescale could be overestimated. Conversely, photometric outliers could shorten this average timescale, making precise estimation challenging.


\section{Conclusions}\label{sec:conclusion}

In this study, we report the first detection of MIR lags in two prototypic IMBHs, NGC 4395 and POX 52, as part of the IMBH-RM project. Our findings are summarized as follows:

\begin{itemize} 
    \item We detected lags of $3.0^{+2.4}_{-1.9}$ days for NGC 4395 and $35.2^{+14.2}_{-11.7}$ days for POX 52 between the optical and WISE W1 bands, based on ICCF and JAVELIN methods, and further supported by PyROA. These results confirm the presence of low-mass black hole activity in both dwarf galaxies.
    
    \item The lag results for these two IMBHs are generally consistent with the extension of the $R_{\mathrm{torus}} - L_{5100}$ relation into the low-luminosity, low-mass regime. However, additional IMBH measurements are needed to establish a robust relation.
    
    \item Using the empirical $R_{\rm torus}-R_{\rm BLR}$ relation, we predict the BLR lags ($\sim$ 4 days) for POX 52, and constrain its BH mass to the range of $10^{5.1 - 5.9}\ M_{\odot}$, thereby generally confirming its IMBH nature through dust RM.
    
    \item We also detected a potential optical inter-band continuum lag of $\lesssim$1 days for POX 52. However, no significant intranight variability was observed, likely due to the longer duty cycle of fast variability in POX 52 compared to NGC 4395. 
\end{itemize}

This work demonstrates that even with very low-cadence IR data, reliable lag measurements can still be obtained if optical data sampling is sufficiently dense and exhibits significant variability. It also provides new insights into the MIR variability of IMBHs and their mass measurements, laying the foundation for future MIR RM of additional low-mass BHs to examine the $R_{\rm torus} - L_{5100}$ relation. 

\bigskip
\noindent We thank the anonymous referee for helpful comments that improved the manuscript. We thank J.W., Lyu, W.J. Liu for useful discussions and J.H., Woo for sharing spectral data. We acknowledges support from the National Key R\&D Program of China No.~2022YFF0503402, 2022YFF0503401, 2023YFA1607903. HXG acknowledges the National Natural Science Foundation of China (NFSC, No. 12473018). MYS acknowledges support from the NSFC (No. 12322303) and the Natural Science Foundation of Fujian Province of China (No. 2022J06002). MFG is supported by the National Science Foundation of China (grant 12473019), the Shanghai Pilot Program for Basic Research-Chinese Academy of Science, Shanghai Branch (JCYJ-SHFY-2021-013), and the National SKA Program of China (Grant No. 2022SKA0120102). MM acknowledges support from the Spanish Ministry of Science and Innovation through the project PID2021-124243NB-C22. This work was partially supported by the program Unidad de Excelencia Maria de Maeztu CEX2020-001058-M. OV acknowledges support from the Astronomical station Vidojevica, funding from the Ministry of Science, Technological Development and Innovation of the Republic of Serbia (contract No.451-03-136/2025-03/200002), by the EC through project BELISSIMA (call FP7-REGPOT-2010-5, No. 265772). OV is partially supported by the Chinese Academy of Sciences (CAS) President's International Fellowship Initiative (PIFI) (grant No. 2024VMB0006). ABK, DI and L\v CP acknowledge funding provided by the University of Belgrade - Faculty of Mathematics (the contract 451-03-136/2025-03/200104) and Astronomical Observatory Belgrade (the contract 451-03-136/2025-03/200002) through the grants by the Ministry of Science, Technological Development and Innovation of the Republic of Serbia. LCH was supported by the National Science Foundation of China (12233001). ZFY is supported by KIPAC Fellowship from Stanford University.

\facility{LCOGT, WISE, ZTF, ATLAS, Gaia}
\software{AstroPy \citep{Astropy2018}}

\bibliography{ms.bib}{}
\bibliographystyle{aasjournal}

\appendix

\section{Short-term MIR variability}\label{app:MIR_daily}
We examined the NEOWISE single-exposure data for NGC 4395 and POX 52 to explore potential short-timescale MIR variability. Although WISE scans the sky every six months, its solar-system-tracking strategy yields multiple exposures of the same region over several days per visit, allowing day-scale variability checks. For NGC 4395, 4 of 22 such visits show potential variabilities at the $\gtrsim$0.1 mag, above the typical scatter seen in nearby reference stars. As shown in Figure~\ref{fig:MIR_daily}, the first visit displays a relatively reliable MIR brightening trend, supported by multiple data points collected over two different days, although no contemporary optical data are available. In contrast, the remaining three visits each contain only a single deviating data point on a separate day, making them more susceptible to being statistical outliers. However, after correcting for the expected MIR delay of 3 days relative to optical band, the variations match the trends seen in the optical light curves. This consistency suggests the variations may be real, consistent with reprocessed emission from a compact dusty torus located a few light-days from the central source.

\setcounter{figure}{0} 
\renewcommand{\thefigure}{A\arabic{figure}}
\begin{figure*}[h!]
\hspace{0.5cm}
    \centering
    \includegraphics[width=0.85\linewidth]{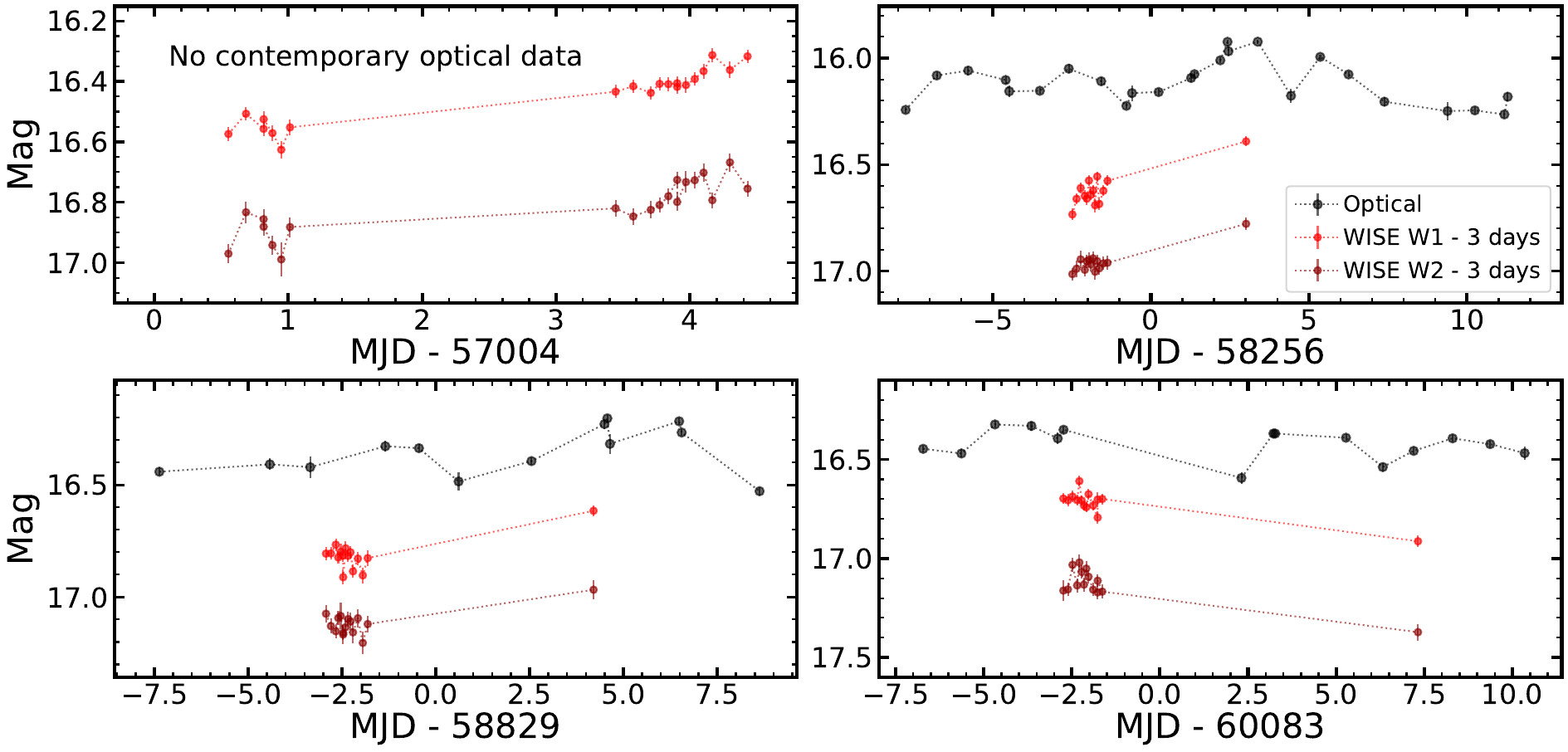}
    \caption{The shifted  MIR Light curves from the single exposures in four visits of NGC 4395 with the contemporary optical light curve. The optical in this figure consists of ATLAS $co$ and ZTF $gr$ bands. The MIR variability shows a significant correlation with the optical light curves, with a time lag of approximately several days, despite the exact time lag can not be precisely constrained due to limited data points.}
    \label{fig:MIR_daily}
\end{figure*}

\section{Lag posteriors and alias effect}\label{app:alias}
Subject to the time sampling and monitoring baseline, all three methods may exhibit multiple peaks in their lag posterior distributions. Therefore, we performed alias removal, i.e., removing false peaks in the lag posteriors if possible, which is particularly necessary when the light curves have significant gaps. For instance, the cadence of WISE data is only once every 180 days, while the seasonal gap of ZTF and ATLAS is around 150 days for POX 52. These gaps can introduce false lag peaks. Figures \ref{fig:NGC4395_post} and \ref{fig:POX52_post} show the lag results for NGC 4395 and POX 52 between the red optical and WISE bands, respectively (see also Table \ref{tab:lag}). The lag measurements for NGC 4395 are dominated by a single peak and consistent among three methods, so we have not applied the alias removal and adopt the ICCF lag as the fiducial value, as it is independent of the model assumption and initial guess.

In POX 52, the seasonal gap ($\sim$120-180 days) introduces a strong alias-induced false peak, as shown in Figure \ref{fig:POX52_post}. In most ICCF posteriors, the  lag posterior is blended with this false peak, making it difficult to remove the aliasing effect. The only exception is the dip segment, where the ICCF result reveals two peaks near 40 days and 95 days. The influence of false peak is more pronounced in the JAVELIN results, where a clear peak appears around 150 days. Given that the false peak is relatively well-separated in JAVELIN posteriors, allowing us to focus on the 0-120 day range (green region) for more reliable lag determination. In the dip segment, JAVELIN shows a clearer separation between two peaks, yielding a $\sim$40-day lag that is broadly consistent with PyROA and with ICCF (after excluding the $\sim$95-day alias), and is considered more reliable. We therefore adopt the JAVELIN (dip) lag as the fiducial value for POX 52, as its false peak is more easily removed compared to ICCF, and it remains consistent with PyROA.

\setcounter{figure}{0} 
\renewcommand{\thefigure}{B\arabic{figure}}
\begin{figure*}[h!]
    \centering
    \includegraphics[width=0.9\linewidth]{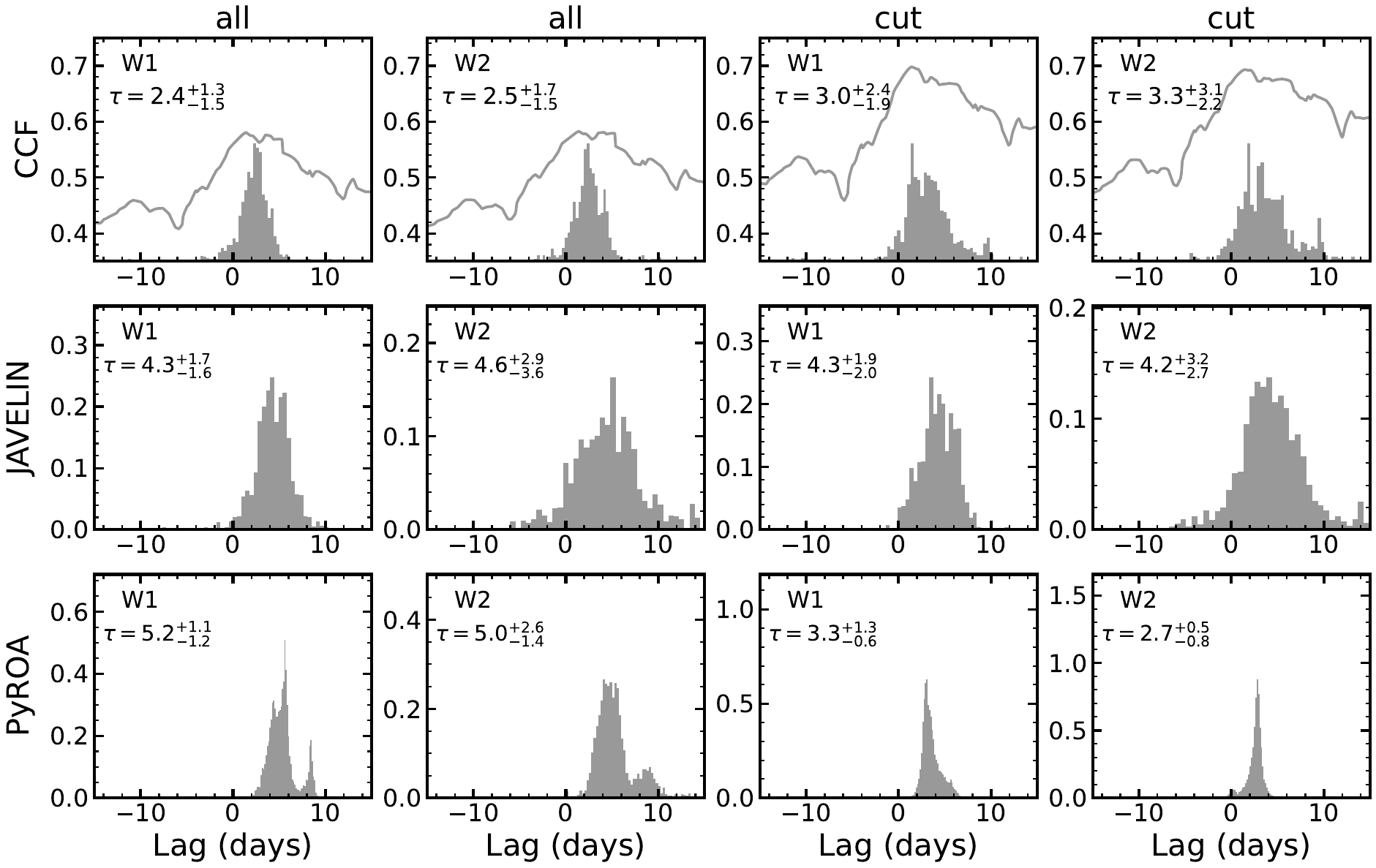}
    \caption{Lag measurements of NGC 4395 between red optical and MIR bands. ICCF, JAVELIN, and PyROA results are shown in the top, middle, and bottom panels, respectively. Each column corresponds to a different lag window and WISE band as listed in Table \ref{tab:lag}.}
    \label{fig:NGC4395_post}
\end{figure*}

\begin{figure*}[h!]
    \centering
    \includegraphics[width=0.9\linewidth]{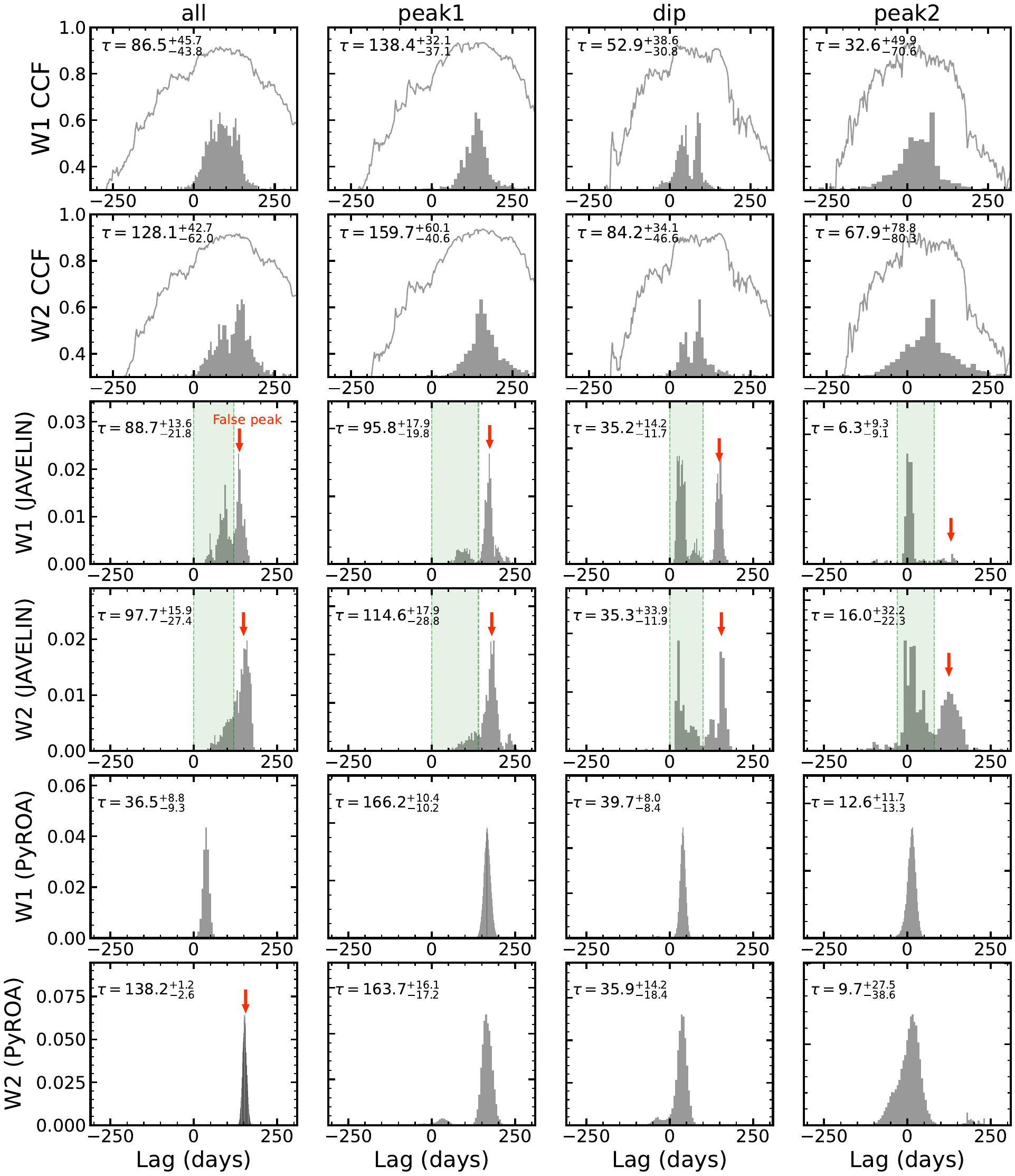}
    \caption{Lag measurements of POX 52 between red optical and MIR bands. The top, middle, and bottom rows show results from ICCF, JAVELIN, and PyROA, respectively. Each column corresponds to a different lag window from Table \ref{tab:lag}. Red arrows mark false peaks caused by seasonal gaps. The green shaded regions ($\sim$0-120 days) are relatively free from aliasing and used for final lag determination.}
    \label{fig:POX52_post}
\end{figure*}

\section{Lag Recovery From Mocked Light Curves}\label{app:mock}
In this section, we perform simulations to evaluate the lag recovery rate for NGC~4395 and POX~52 under current observational conditions. Following \citet{Guo21} and \citet{U22}, we first fit the observed optical light curves using the DRW model, obtaining parameters of $\tau=37$~days, $\sigma=0.09$ mJy for NGC~4395, and $\tau=193$ days, $\sigma=0.06$ mJy for POX~52. We then generate mock optical light curves 1000 times longer than the observations and select 500 segments whose variability amplitudes differ from the observed ones by less than 20\%. To produce corresponding MIR light curves, the selected optical segments are shifted by the input lags (3 days for NGC~4395 and 35 days for POX~52) and smoothed using the same lag time as the smoothing window, then resampled to match the observed cadence. Gaussian noise is added based on observed flux uncertainties. Finally, we measure the lag by computing the CCF with \texttt{PyI$^2$CCF} (Section\ref{sec:measure_lag}).

Among the 500 simulations of NGC~4395, 372 exhibit strong cross-correlation with \(r_{\rm max} > 0.6\). The corresponding time lags and an example set of mock light curves are shown in Figure~\ref{fig:mock_NGC4395}. We define a successful recovery as being within \(1\sigma\) of the input lag, which is a more strict criterion compared to the \(3\sigma\) definition adopted in previous work \citep[e.g.,][]{Li19}. This choice is motivated by the relatively large uncertainties in our time lag measurements, which could encompass all recovered lags within the $3\sigma$ range. Under this criterion, about 50\% of the lags are recovered, with a median lag with $1\sigma$ uncertainty of \(2.3^{+1.5}_{-1.6}\) days for the sample with \(r_{\rm max} > 0.6\). This value is slightly lower than the input lag of 3 days, likely due to sampling and smoothing effects in the MIR light curve generation \citep{Yu20}, but remains broadly consistent given the scatter. 

The simulation of POX~52 is similar to that of NGC~4395, except that we fix the long-term variability trend (as described in Section~\ref{sec:micro_var}), since the two-peak structure is not easily reproduced by the DRW model. We also perform ICCF measurements in each segment. All simulations show strong correlations with \(r_{\rm max} > 0.8\) under the fixed long-term variability trend. The \(1\sigma\) recovery rates across the four segments are all within \(40\%-60\%\), with median values comparable to the input time lag. The time lags measured in the peak1 and peak2 segments exhibit significantly larger scatter, whereas the results from the full dataset and the dip segment are relatively stable. We define an outlier as a measurement with a deviation from the input time lag exceeding 100\%, i.e., more than 35 days. Based on this criterion, the outlier rates for the all, peak1, dip, and peak2 segments are 13\%, 19\%, 8\%, and 26\%, respectively. Therefore, the dip segment performs slightly better, with reduced scatter and fewer outliers, suggesting that this may be the least affected by seasonal gaps in the observed light curves.

\setcounter{figure}{0} 
\renewcommand{\thefigure}{C\arabic{figure}}
\begin{figure*}[h]
    \centering
    \includegraphics[width=0.8\linewidth]{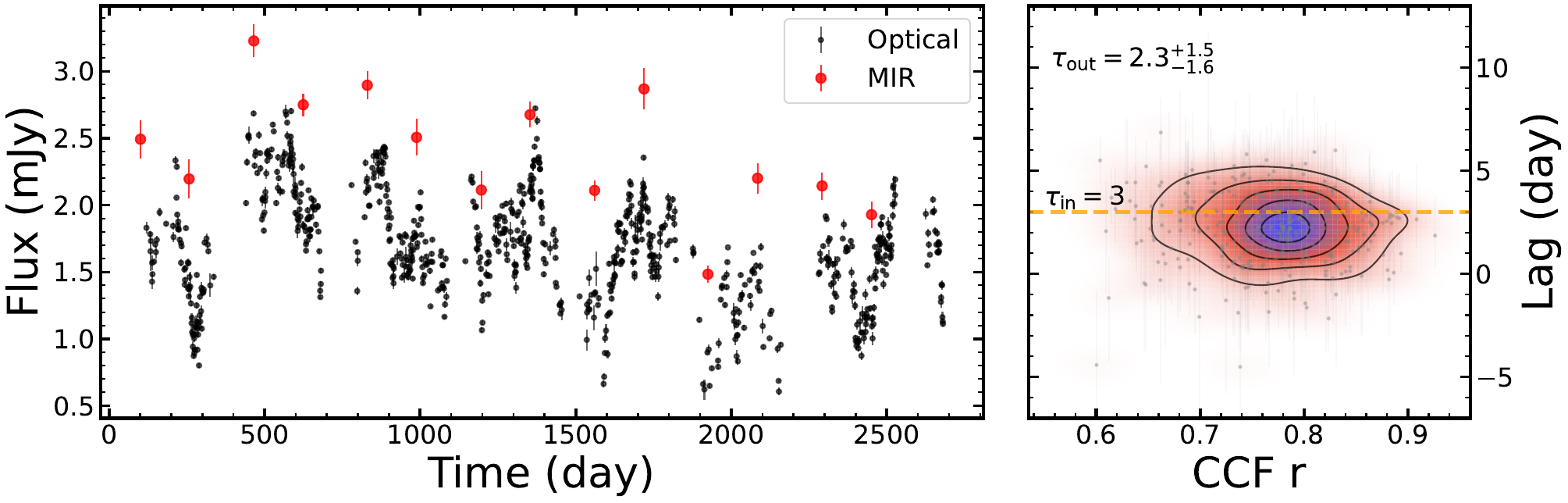}
    \caption{Left panel: an example of the mocked light curves for NGC 4395. Right panel: the measured time lags as a function of CCF $r_{\rm max}$ of the mocked light curves ($r_{\rm max}>0.6$). The contours represent the density distribution, ranging from high (blue) to low (red). The horizontal orange line indicates the input time lag of 3 days.}
    \label{fig:mock_NGC4395}
\end{figure*}

\begin{figure*}[h]
    \centering
    \includegraphics[width=0.8\linewidth]{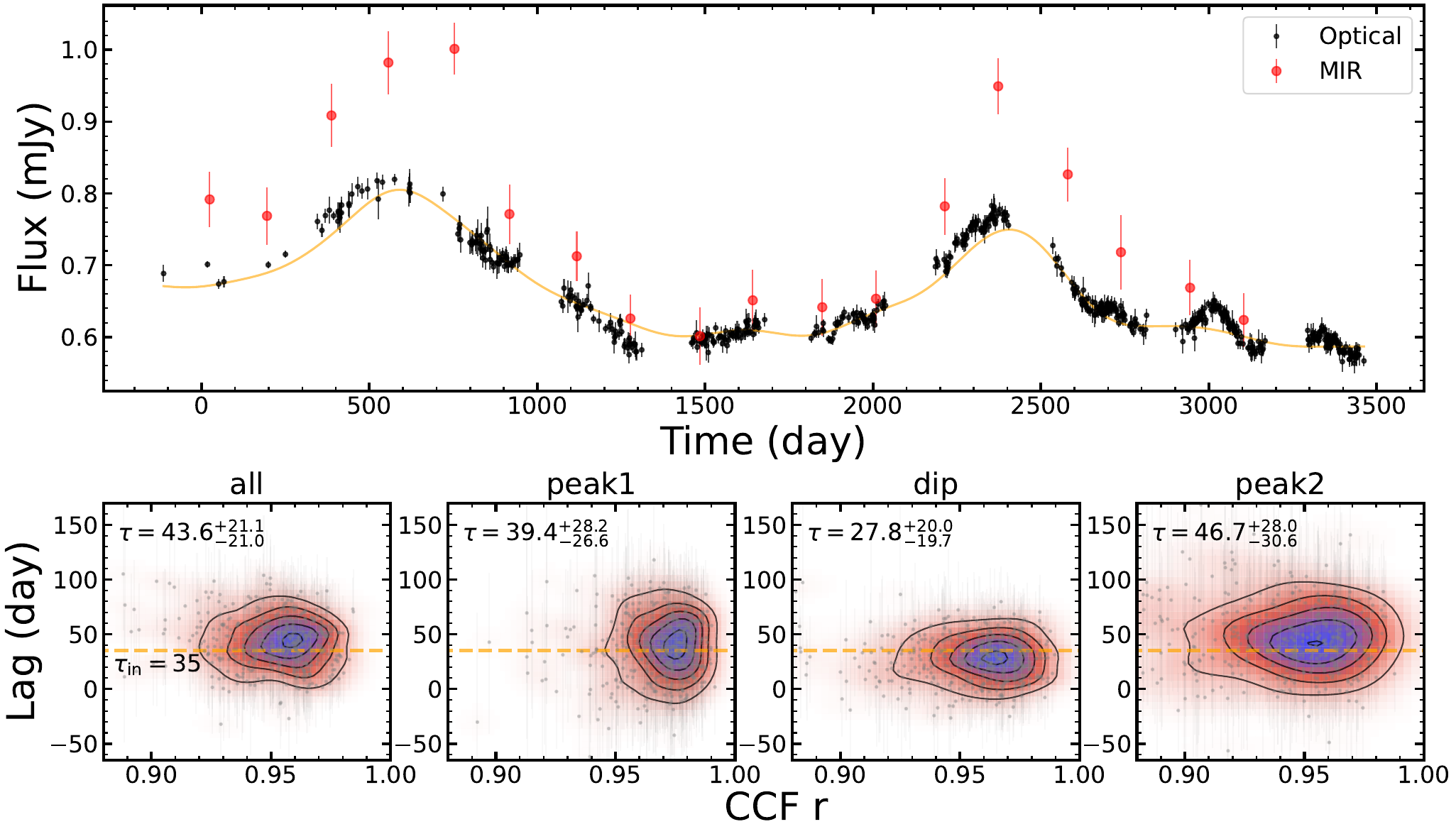}
    \caption{Upper panel: an example of the mocked light curves for POX 52. The solid orange line denotes the fixed long-term variability. Bottom: the measured time lags as a function of CCF $r_{\rm max}$ of the mocked light curves in different segments. The contours represent the density distribution, ranging from high (blue) to low (red). The horizontal orange line indicates the input time lag of 35 days.}
    \label{fig:mock_POX52}
\end{figure*}

\end{CJK}
\end{document}